\documentclass[prc,twocolumn,floatfix,showpacs,superscriptaddress]{revtex4}
\usepackage{amssymb}
\usepackage[centertags]{amsmath}
\usepackage{txfonts}
\usepackage{bm}
\usepackage{xcolor}
\usepackage{epsfig}
\usepackage{graphicx,graphics}
\usepackage{float}
\usepackage[colorlinks,citecolor=blue,linkcolor=blue]{hyperref}

\newcommand{\ovl}[1]{\overline{#1}}

\begin{document}
\title{ Multiplicity fluctuation and correlation of identified baryons in quark combination model}

\author{Jun Song}\email{songjun2011@jnxy.edu.cn} 
\affiliation{Department of Physics, Jining University, Shandong 273155, China}

\author{Hai-hong Li}
\affiliation{School of Physics and Engineering, Qufu Normal University, Shandong 273165, China}
\affiliation{Department of Physics, Jining University, Shandong 273155, China}

\author{Rui-qin Wang}
\affiliation{School of Physics and Engineering, Qufu Normal University, Shandong 273165, China}

\author{Feng-lan Shao}\email{shaofl@mail.sdu.edu.cn} 
\affiliation{School of Physics and Engineering, Qufu Normal University, Shandong 273165, China}

\begin{abstract}
	The dynamical multiplicity fluctuations and correlations of identified baryons and antibaryons produced by the hadronization of the bulk quark system are systematically studied in quark combination model. Starting from the most basic dynamics of the quark combination which is necessary for multiplicity study, we analyze moments (variance, skewness and kurtosis) of inclusive multiplicity distributions of identified baryons, two-baryon multiplicity correlations, and baryon-antibaryon multiplicity correlations after the hadronization of quark system with given quark number and antiquark number. We obtain a series of interesting results, e.g.,~binomial behavior of multiplicity moments, coincide flavor dependent two-baryon correlation and universal baryon-antibaryon correlation, which can be regarded as general features of the quark combination. We further take into account correlations and fluctuations of quark numbers before hadronization and study their influence on multiple production of baryons and antibaryons. We find that quark number fluctuations and flavor conservation lead to a series of important results such as the negative $p\bar{\Omega}^{+}$ multiplicity correlation and universal two-baryon correlations. We also study the influence of resonance decays in order to compare our results with future experimental data in ultra-relativistic heavy ion collisions at LHC. 
\end{abstract}
\maketitle
\section{introduction}
In ultra-relativistic heavy ion collisions, a new state of the matter --- Quark Gluon Plasma (QGP) is created at the early stage of collisions. The produced QGP expands, cools and changes into a hadronic system at a critical energy density \cite{qgp2004}. Because of the non-perturbative difficulty of quantum chromodynamics, the transition from QGP to hadrons (i.e.,~hadronization) can only be described currently by phenomenological models such as statistical hadronization models \cite{becattini02,shm03} and quark (re-)combination/coalescence models \cite{alcor2000,Fries22003prl,ckm03,hwa04,nonaka05,co2006PRC,sdqcm}. These models have been tested against the available experimental data of hadronic yields, momentum spectra and flows. 

Dynamical correlations and fluctuations of multi-hadron production carry more sophisticated hadronization dynamics.  They are quantified by various covariances and moments on multiplicities or momenta of identified hadrons, and are measured in experiments via event-by-event method. Their studies can further test those existing phenomenological models of hadron production at hadronization and gain deep insights on dynamics of realistic hadronization process. We can also obtain the information of the correlations and fluctuations of quarks and antiquarks just before hadronization by studying their projection on hadronic observables. On the other hand, study of identified hadrons is also helpful for the investigation of correlations and fluctuations of conservative charges which is a hot topic both in experimental and theoretical studies recently \cite{star12Moments,kochFluct,StephanovFluct,Asakawa09,Karsch11}. There one should know how the conservative charges populate in various identified hadrons, which depends on their coherent abundances and thus is directly related to their multiple production dynamics at hadronization.

In the past few years, only data on fluctuations of the pion, kaon and proton are reported \cite{Na492009,Na492011,star2009,star2015} and the available theoretical studies are mainly of them usually based on statistical model \cite{koch08,fjh0912,tawfik,Gorenstein2009,Konchakovski2009,koch2010, torr2007}. With the improvement of statistics and experimental measurement precision, observation of more hadron species such as $\Lambda$, $\Xi^{-}$ and $\Omega^{-}$ can be expected in the near future. Therefore, the corresponding theoretical predictions by different hadron production models are necessary, which are used to guide the experimental data analysis, reveal the underlying dynamics of the observation and test these models.

In this paper, we study the multiplicity fluctuations and correlations of various identified baryons and antibaryons produced directly by hadronization. We focus on the $J^P = \frac{1}{2}^{+}$ and $\frac{3}{2}^{+}$ baryons in flavor SU(3) ground state with particular emphasis on various strange baryons.  There are obvious advantages in measuring these baryons: (1) baryon is a sensitive probe of hadron production mechanism at hadronization. (2) the rapidity shift in baryon productions and resonance decays is small, which is suitable to experimental observation at finite rapidity window size. 

We use the quark combination mechanism (QCM) to describe the production of hadrons at quark system hadronization. QCM has been used to reproduce lots of low and intermediate transverse momentum data at RHIC and LHC, in particular the data of yields and rapidity distributions \cite{sdqcm,SJ13,wrq12,wrq15}. The related entropy and pion production issues have been extensively addressed in literatures \cite{biro1999,biro2007entropy,fries08review,sj10s}. Explaining fluctuations and/or correlations of hadron production is very intuitive in QCM. When a quark hadronizes, it can come into either a baryon or a meson, which leads to the fluctuation of global baryon multiplicity; it can come into either a specific baryon (e.g.,~a proton for a $u$ quark hadronization) or another specific baryon (e.g.,~ a $\Delta^{+}$), which leads to the multiplicity fluctuations of proton and $\Delta^{+}$ and also an anti-correlation between two baryons. In addition, correlations and fluctuations of quarks and antiquarks will pass to hadrons after hadronization.

Concretely, we calculate various moments of inclusive multiplicity distributions of baryons, e.g.,~variance, skewness and kurtosis, the correlations between two baryons and correlations between baryons and antibaryons.  We analyze the dominant dynamics among these correlations and fluctuations and give predictions of QCM which can be tested by the future experimental data. This paper mainly discusses baryon production at zero baryon number density at LHC, and the extension to RHIC energies and meson sector is the goal of the future work.

The paper is organized as follows. In section II, we introduce a working model which includes the necessary dynamics of QCM for multiplicity study and discuss the dynamical sources of the multiplicity correlations and fluctuations in baryon production.  In Section III, we study multiplicity fluctuations and correlations of baryons and antibaryons which are produced from the quark system with the given numbers of quarks and antiquarks. In Section IV, we take into account fluctuations and correlations of quark numbers before hadronization to study their influence on baryon and antibaryon production. In Section V, we further take into account effects of resonance decays.  Summary and discussion are given in Sec VI. 

\section{a working model}

Due to the difficulty of non-perturbative QCD, a widely-accepted theoretical framework of QCM is not established so far which can self-consistently describe the whole picture of hadronization dynamics. In this paper, we need a working model which includes the necessary dynamics of QCM for multiplicity study and obtain correlations and fluctuations of the produced hadrons.  We will present the assumptions and/or inputs explicitly whenever necessary and make the study as independent of the particular model as possible. Because there are no relevant works in literatures, the purpose of this paper is to focus on results of the most basic QCM dynamics which will serve as a preliminary test of the model using the future experimental data and a baseline for the sophisticated hadronization dynamics. 

We consider a system consisting of various quarks and antiquarks with constituent masses, corresponding to the ``dressed'' quarks and antiquarks in non-perturbative QCD regime. We denote the number of quarks of flavor $q_i$ in the system by $N_{q_i}$ and that of antiquarks by $N_{\bar{q}_i}$. Three flavors, up, down and strange, are considered in this paper.  As the system hadronizes, these quarks and antiquarks combine with each other to form color singlet hadrons.  Finally, the system produces, in an event, various hadrons with numbers $\{ N_{h_i}\}$ where $i =\pi, K, \rho, K^*, ...., p, \Lambda, \Xi, \Omega^{-}$ up to all included hadron species. Here, we consider only the ground state $J^P = 0^-$ and $1^{-}$ mesons and $J^P = \frac{1}{2}^{+}$ and $\frac{3}{2}^{+}$ baryons in flavor SU(3) group.
The numbers of these hadrons are varied event-by-event around their average values and follow a certain distribution $\mathcal{P}(\{ N_{h_i}\};\{ N_{q_j}, N_{\bar{q}_j}\})$ which is governed by hadronization dynamics.

The precise form of $\mathcal{P}(\{ N_{h_i}\};\{ N_{q_j}, N_{\bar{q}_j}\})$ depends on the full knowledge of hadronization dynamics. On all the ``on market'' QCM models, few ones can give their specific solutions of $\mathcal{P}$. In addition, high dimensionality feature of $\mathcal{P}$ makes the analytic solution quite difficult to get. In this paper, we generalize the quark combination simulation in SDQCM \cite{sdqcm} to focus only on multiplicity properties of the produced hadrons and obtain the $\mathcal{P}(\{ N_{h_i}\};\{ N_{q_j}, N_{\bar{q}_j}\})$, considering that this model has reproduced lots of experimental data of multiplicities of various hadrons in relativistic heavy ion collisions at different energies \cite{sdqcm,SJ13,wrq12,wrq15}.  

The main idea of the quark combination simulation in SDQCM is as follows: (1) assign all quarks and antiquarks in system into an abstract one-dimensional sequence. The relative distance between any two quarks and/or antiquarks in the sequence represents their map in realistic phase space. (2) combine these quarks and antiquarks in the sequence into hadrons according to a quark combination rule (QCR). A schematic example is as follows
\begin{eqnarray}
\label{example} &&q_1\ovl{q}_2\ovl{q}_3\ovl{q}_4\ovl{q}_5q_6
\ovl{q}_7q_8q_9q_{10}\ovl{q}_{11}q_{12}q_{13}q_{14}\ovl{q}_{15}
q_{16}q_{17}\ovl{q}_{18}\ovl{q}_{19}\ovl{q}_{20}\nonumber\\
&&\stackrel{QCR}{\rightarrow} M(q_1\ovl{q}_2)\;\ovl{B}(\ovl{q}_3\ovl{q}_4\ovl{q}_5)\;
	M(q_6\ovl{q}_7)\;M(q_8 \ovl{q}_{11})\;B(q_9 q_{10} q_{12}) \;\nonumber\\
&&\hspace{10pt} M(q_{13}\ovl{q}_{15})\;B(q_{14}q_{16}q_{17})\;
\ovl{B}(\ovl{q}_{18}\ovl{q}_{19}\ovl{q}_{20}). 
\end{eqnarray}

QCR depends on the combination dynamics. As shown directly by the above example, QCR should firstly satisfy two basic dynamics:~ (1) baryon formation is by the combination of three quarks which are close with each other in phase space and meson by a quark and an antiquark. Therefore, neighboring or next-neighboring quark combination in the sequence is needed; (2) after hadronization, there are no free quarks and antiquarks left. 

Considering the fact that the produced baryons are much less than mesons after hadronization, the key content of QCR is how to describe the production of baryons relative to that of mesons for a given quark configuration. We adopt the following procedure.  For the local quark populations such as $q\bar{q}$ and $qq\bar{q}$, we can assign $q\bar{q}\rightarrow M$ and $qq\bar{q}\rightarrow M + q$ with relative probability 1. When the case of possible baryon production $qqq$ occurs, we give a probability or conditional criterion.  If the nearest neighbor of $qqq$ is still a quark, the opportunity of baryon formation should be significantly increased, and we can assign $qqqq\rightarrow B + q$ with relative probability 1. On the contrary, if the nearest neighbor of $qqq$ is an antiquark $\bar{q}$, then this $\bar{q}$ can have the chance of capturing one $q$ to form a meson and two quarks are left to combine with other quarks and antiquarks. We denote the probability of this channel by $P_{qqq\bar{q}\rightarrow M + qq}\equiv P_0$. The baryon formation probability in $qqq\bar{q}$ configuration is then $P_1 \equiv P_{qqq\bar{q}\rightarrow B + \bar{q}} =  1- P_0$. 

A naive analysis gives $P_0/P_1 \sim (3\times \frac{1}{9}) / (1 \times \frac{1}{27})=9$ where the factor 3 is the number of the possible combinations for meson formation in $qqq\bar{q}$ configuration and factor 1 for baryon formation. Factor $\frac{1}{9}$ and $\frac{1}{27}$ are the color weights of forming color singlet meson and baryon in the stochastically colored quark combination, respectively. Therefore, baryon formation probability $P_1$ in $qqq\bar{q}$ case should be a small value $\sim 0.1$. In practice, a value of about 0.04 for $P_1$ can well explain the observed baryon yields in relativistic heavy ion collisions. 

Above consideration in baryon formation is one kind of non-isolation approximation for the quark combination process, i.e.,~ baryon formation is non-trivially influenced by the environment (the surrounding quarks and antiquarks).
It is different from those (re)combination/coalescence models which are popular at early RHIC experiments \cite{Fries22003prl,ckm03,hwa04,nonaka05,co2006PRC}. They apply the sudden hadronization (i.e.,~isolation) approximation for the combination probability by the overlap between quark wave function and the hadron they form. 

The remaining quarks and antiquarks in $qq\bar{q}\rightarrow M +q$, $qqq\bar{q}\rightarrow M + qq$ and $qqq\bar{q}\rightarrow B + \bar{q}$ processes will subsequently combine with following quarks and antiquarks in the sequence to form hadrons until at last all quarks and antiquarks are combined into hadrons. This procedure reflects, to a certain extent, the spread of the hadronization in space-time.

Another point of QCR is the order of the combination.  As long as the quark number is large, different orders such as from left to right, from right to left, and from middle to sides are equivalent and give the same result. 

Based on the above discussions, we give the following combination algorithm for the hadronization of quark system:
\begin{description}
\item (i) start from the first parton ($q$ or $\bar{q}$ ) in the sequence.
\item (ii) if the first and second partons are either $\bar{q}q$ or $q\bar{q}$, they combine into a meson and are removed from the sequence, then go back to (i); if the first two are $qq$ or $\bar{q}\bar{q}$, and then go to the next.
\item (iii) look at the third parton, if three partons are $qq\bar{q}$ or $\bar{q}\bar{q}q$, and the first and third partons combine into a meson and are removed from the sequence, and then go back to (i); if three partons are $qqq$ or $\bar{q}\bar{q}\bar{q}$ then go to the next.
\item (iv) look at the fourth parton , if four partons are $qqqq$ or $\bar{q}\bar{q}\bar{q}\bar{q}$, the first three partons combine into a baryon or an antibaryon and are removed from the sequence, and then go back to (i); if four partons are $qqq\bar{q}$ or $\bar{q}\bar{q}\bar{q}q$, there are two choices: (a) the first and fourth partons combine into a meson with probability $P_0$ and are removed from the sequence, and  then go back to (ii); (b) the first three partons combine into a baryon or an antibaryon with probability $P_1$ and are removed from the sequence, and then go back to (i).
\end{description}

Above algorithm does not differentiate quark flavors in consideration of the flavor blind of strong interactions. 
Compared with the combination rule in Ref.~\cite{sdqcm}, this algorithm addresses more explicitly the baryon production by the addition of step (iv) to better tune baryon meson production competition. In essence, it can be regarded as the generalization of the combination rule in Ref.~\cite{sdqcm} in multiplicity description of the produced baryons. 

For a given $q_1 \bar{q}_2$ which is known to form a meson by the above combination algorithm, it can form either a $J^{p}=1^-$ vector (V) meson or a $J^p = 0^-$ pseudo scalar (PS) meson. Similarly, a $q_1 q_2 q_3$ (except for three identical $qqq$ case) can form either a $J^{p}=(\frac{1}{2})^+$ baryon or a $J^{p}=(\frac{3}{2})^+$ baryon. Following previous works \cite{sdqcm,wrq12}, we use the parameter $R_{V/P}$ to denote the relative production ratio of vector mesons to pseudoscalar mesons and $R_{O/D}$ the ratio of octet baryons to decuplet baryons. Then we get the branch ratio of each hadronization channel for a $q_1 \bar{q}_2$ combination
\begin{equation}
   C_{M_j} =  \left\{
	\begin{array}{ll}
		   {1}/{(1+R_{V/P})}~~~~~~~~   \textrm{for } J^P=0^-  \textrm{ mesons},  \\
			{R_{V/P}}/{(1+R_{V/P})}~~~      \textrm{for } J^P=1^-  \textrm{ mesons},
	\end{array} \right.  \nonumber
\end{equation}
and for a $q_1 q_2 q_3$ combination  
\begin{equation}
C_{B_j} =  \left\{
    \begin{array}{ll}
	      {R_{O/D}}/{(1+R_{O/D})}~~~~   \textrm{for } J^P=({1}/{2})^+  \textrm{ baryons},  \\
			{1}/{(1+R_{O/D})}~~~~~~~~~~         \textrm{for } J^P=({3}/{2})^+  \textrm{ baryons}.
	 \end{array} \right. \nonumber
\end{equation}

As did in previous works, we can apply the above combination algorithm to relativistic heavy ion collisions by considering some properties of the produced quark system. It is observed that (1) the longitudinal expansion is predominant both in momentum space and in spatial space; (2) the longitudinal velocity of quarks is closely correlated to their spatial position; (3) the rapidity density of quark numbers is very large and is relatively slowly varied. Therefore, we can sort all quarks and antiquarks according to their rapidities into an one-dimensional sequence, and then combine neighboring quarks and antiquarks into hadrons. In the transverse direction, transverse momentum ($p_T$) distribution of quarks is exponential decreased. Therefore, one can not directly combine neighboring quarks because their relative intervals $\Delta p_T$ exponentially increase with $p_T$ of quarks/antiquarks. So we use the statistical combination approach, i.e.,~the $p_{T}$ distribution of hadron is the convolution of quark $p_{T}$ distributions and combination kernel, where the combination kernel is mainly dependent on $\Delta p_T$ between two quarks/antiquarks. It is thus similar to those inclusive recombination/coalescence approaches using the hadron wave function \cite{Fries22003prl,ckm03,hwa04,nonaka05,co2006PRC}. But our model is different from those inclusive methods in the proper treatment of unitarity in hadronization and the ability of well explanation of hadronic yield and longitudinal rapidity distributions observed in relativistic heavy ion collisions \cite{sdqcm,SJ13,wrq12,wrq15}. 

Let us summarize the origin of correlations and fluctuations of the produced baryons and antibaryons. First, local $qqq$ aggregation in phase space is stochastic for the system consisting of free quarks and antiquarks. Second, the $qqq\rightarrow B$ process is probabilistic under the noise surrounding (i.e.,~stochastic populated quarks and antiquarks in neighbourhood). Together with the branch ratio of a given $q_1 q_2 q_3$ to a specific hadron state, they lead to multiplicity fluctuations of the produced identified baryons. The conservation of baryon number in quark combination process constrains the global production of baryons and antibaryons and also the production of identified baryons and their anti-particles. The production correlation between two baryons mainly comes from a so-called ``exclusion'' effect, i.e.,~ once a quark enters into a $B_{i}$ at hadronization it is consumed and therefore can not be recombined into $B_{j}$.  These effects lead to a nontrivial and complex multi-hadron multiplicity distribution $\mathcal{P}(\{ N_{h_i}\};\{ N_{q_j}, N_{\bar{q}_j}\})$.

\section{baryon production from a given quark system}
In this section, we study fluctuations and correlations of baryons and antibaryons which are produced from the quark system with the given number of quarks and antiquarks. This enables us to learn more clearly the properties of baryon production from the quark combination process itself. Analytical results of various moments (mean, variance, skewness, kurtosis) of the inclusive multiplicity distributions of identified baryons are given firstly, according to the basic dynamics of the quark combination discussed in previous section. Then two-baryon multiplicity correlations, baryon-antibaryon correlations and multi-baryon multiplicity correlations are studied systematically.

\subsection{ moments of multiplicity distributions of baryons}
\begin{figure}[!htbp]
\centering
  \includegraphics[width=\linewidth]{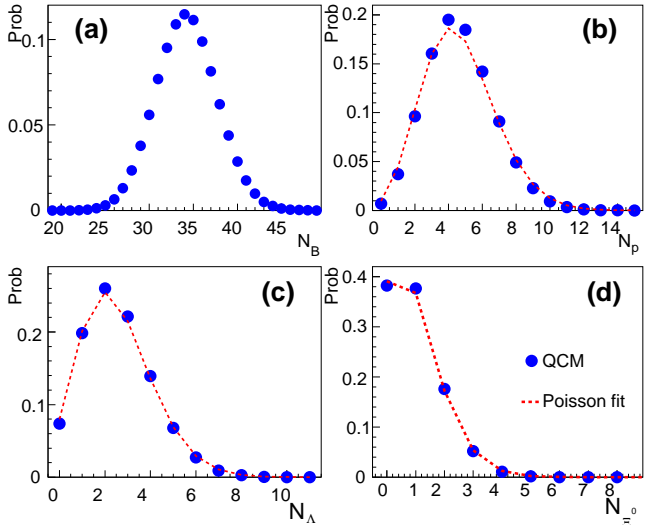}\\
  \caption{(Color online) Normalized multiplicity distribution of total baryons (a) and  those of identified baryons $p$ (b), $\Lambda$ (c), $\Xi^{0}$ (d) produced by hadronization of a quark system with $N_q = N_{\bar{q}} = 500$. Here, the relative ratios of different quark flavors are set to be $N_{u}:N_{d}:N_{s}=1:1:0.43$. Symbols are numerical results of QCM algorithm in Sec.~II and the dashed line is Poisson distribution.}\label{fig1}
\end{figure}

Firstly, we discuss properties of inclusive multiplicity distributions of various identified baryons calculated from the above combination algorithm.
As a demonstration, Fig.~\ref{fig1} shows multiplicity distribution of total baryons and those of identified $p$, $\Lambda$, $\Xi^{0}$, as the quark system with $N_q = N_{\bar{q}} = 500$ hadronizes. Here, the relative ratios of different quark flavors are set to be $N_{u}:N_{d}:N_{s}=1:1:0.43$. We see that the distribution of total baryons is close to the Gaussian distribution while those of identified baryons are close to Poisson distribution to a certain extent. In the following text, we study the production property of these identified baryons by analyzing moments of their multiplicity distributions.

 For the average multiplicity of identified baryons, 
\begin{equation}
  \overline{N}_{B_j} =\sum_{\{ N_{h_i}\}} {N}_{B_j} \  \mathcal{P}(\{ N_{h_i}\};\{ N_{q_i}, N_{\bar{q}_i}\}),
  \label{avnbi}
\end{equation}
we have obtained the empirical solution in previous studies \cite{sdqcm,wrq12}
\begin{equation}
	\overline{N}_{B_j} = P_{B_j}\ \overline{N}_{B}
	\label{AveBi}
\end{equation}
where $\overline{N}_B=\sum_{i} \overline{N}_{B_i}$ is the average number of total baryons and $P_{B_j}$ denotes the production weight of $B_j$ in all baryons. $P_{B_j}$ can be decomposed to $C_{B_j} P_{q_1 q_2 q_3,B} $ where $P_{q_1 q_2 q_3,B}$ is the probability that, as a baryon is known to be produced, the flavor content of this baryon is $q_1 q_2 q_3$. Considering that every $q_1$, $q_2$ and $q_3$ in the system can have the chance of entering into $B_j$ at hadronization, we get $P_{q_1 q_2 q_3,B}= N^{(q)}_{B_j}/N_{qqq}$. $N_{qqq}=N_q(N_q -1)(N_q-2)$ is the possible total number of three quark combinations where $N_{q} = \sum_f N_f$ is total quark number in system. $N^{(q)}_{B_j} = N_{iter} \prod_{f} \prod_{i=1}^{n_{f,B_j}} (N_f -i +1)$ is the possible number of $q_1 q_2 q_3$ combinations where $n_{f,B_j}$ is the number of valance quark $f$ contained in hadron $B_j$. Here index $f$ runs over all quark flavors. $N_{iter}$ is the iteration factor taking to be 1, 3, and 6 for the case of three identical flavor, two different flavors and three different flavors contained in a baryon, respectively.

We have used Eq.~(\ref{AveBi}) to reproduce the experimental data of yields and yield ratios of various identified baryons in relativistic heavy ion collisions at different collision energies\cite{sdqcm,SJ13,wrq12,wrq15}. For the detailed discussions of the average yield formula of identified baryons as well as those of antibaryons we refer readers Refs. \cite{SJ13,wrq12,wrq15}. 
We argue that just based on these well performance of combination algorithm in Sec. II on the event-average yields, we make further test in fluctuations and correlations in this paper.

We further study the variance, skewness and kurtosis of multiplicity distribution for various identified baryons. Their definitions are 
\begin{equation}
\begin{split}
\overline{\sigma}^2_{B_j} &=\overline{{\delta N_{B_j}}^2 } = \overline{({N}_{B_j}-\overline{N}_{B_j})^2} \\
		&=\sum_{\{ N_{h_i}\}} ({N}_{B_j}-\overline{N}_{B_j})^2 \  \mathcal{P}(\{ N_{h_i}\};\{ N_{q_j}, N_{\bar{q}_j}\}),
\end{split}
\end{equation}
and similarly
\begin{equation}
\overline{S}_{B_j} = \frac{ \overline{ {\delta N_{B_j} }^3 } }{ \overline{\sigma}^3_{B_j} } \hspace{30pt}
\overline{K}_{B_j} = \frac{ \overline{ {\delta N_{B_j} }^4 } }{ \overline{\sigma}^4_{B_j} } - 3.
\end{equation}
Note that we always use the superscript \emph{overline} to denote the average hadronic quantities by hadronization of a given quark system.

\begin{figure*}[!htbp]
  \includegraphics[width=\linewidth]{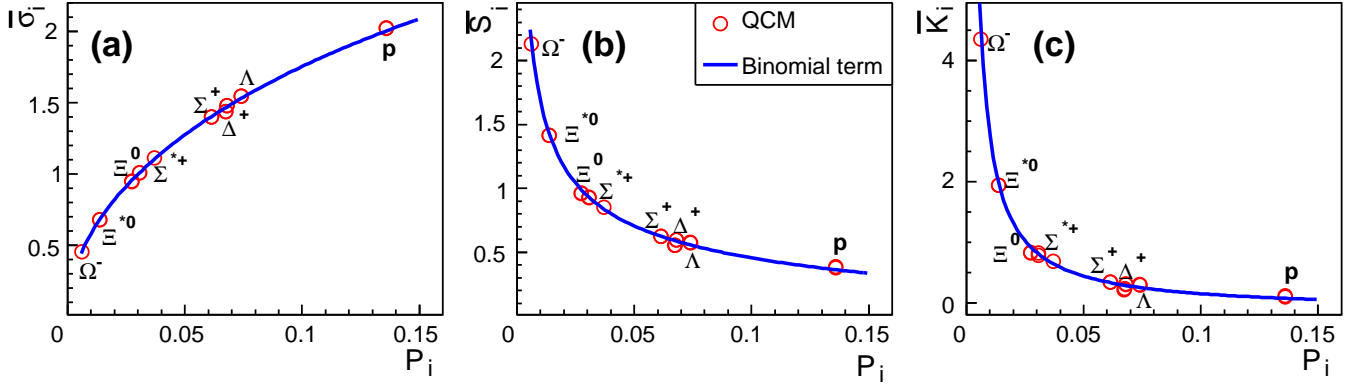}\\
  \caption{(Color online)  The square root of variance, skewness and kurtosis of the multiplicity distributions of various identified baryons with respect to their production weights $P_{i}=\overline{N}_{B_i}/\overline{N}_B$. The size of quark system before hadronization is chosen to be $N_q = N_{\bar{q}} = 500$ and the relative ratios of different quark flavors are set to be $N_{u}:N_{d}:N_{s}=1:1:0.43$. Symbols are full results and lines are leading terms of full results which have the form of binomial distribution with parameter ($\overline{N}_B, P_{i}$).}\label{idbmz0}
  \label{fig2}
\end{figure*}

To analyze their properties, we have to consider joint production of multi-baryons. Taking variance for example, two-$B_j$ pair production is given by
\begin{equation}
	\overline{N_{B_j}(N_{B_j}-1)}= P_{2B_j} \ \overline{N_{B}(N_{B}-1)}, 
\end{equation}
where the production probability of two-$B_j$ pair can be evaluated by $ P_{2B_j}=C^2_{B_j} {N^{(q)}_{2B_j}}/{N_{6q}}$ with the number of six-quark cluster possible for two-$B_j$ pair production $N^{(q)}_{2B_j} = N^{2}_{iter} \prod_{f} \prod_{i=1}^{2 n_{f,B_j}} (N_f -i +1)$ and that for any two-baryon pair production $N_{6q}=\prod_{i=1}^{6}(N_q -i +1)$. Rewriting $P_{2B_j} = P^{2}_{B_j}(1-A_1)$, we finally have
\begin{equation}
	\overline{\sigma}^2_{B_j} = \overline{N}_{B} P_{B_j}(1-P_{B_j}) + P_{B_j}^{2} \Big[ (1-A_1) \overline{\sigma}^2_{B} - A_1  \overline{N}_{B}(\overline{N}_{B}-1) \Big] 
\label{iniSigBi}
\end{equation}
where the first term in the right hand side of the equation is the dominant part.
Similarly, we have 
\begin{widetext}
\begin{equation}
\begin{split}
	\overline{S}_{B_j} = \frac{1}{\overline{\sigma}^3_{B_j}}\Bigg\{ & \overline{N}_{B} P_{B_j}(1-P_{B_j})(1-2 P_{B_j}) 
	+ 3 P_{B_j}^{2} \Big[ (1-A_1) \overline{\sigma}^2_{B} - A_1  \overline{N}_{B}(\overline{N}_{B}-1) \Big] \\
&+ P_{B_j}^{3} \Big[ (1-A_2)\overline{S}_{B} \overline{\sigma}_{B}^3 + 3(A_1 -A_2)\overline{N}_B \overline{\sigma}^2_{B} - 3(1 -A_2)\overline{\sigma}^2_{B}+\overline{N}_{B}(\overline{N}_{B}-1)\big[ (3A_1 -A_2)\overline{N}_{B}+2A_2\big] \Big] \Bigg\}, 
\end{split}
\end{equation}
and 
\begin{equation}
\begin{split}
	\overline{K}_{B_j} +3 = \frac{1}{\overline{\sigma}^4_{B_j}}\Bigg\{ & \overline{N}_{B} P_{B_j}(1-P_{B_j}) \Big[1- 6P_{B_j} (1- P_{B_j}) + 3  \overline{N}_{B} P_{B_j}(1-P_{B_j}) \Big] + 7 P_{B_j}^{2} \Big[ (1-A_1) \overline{\sigma}^2_{B} - A_1  \overline{N}_{B}(\overline{N}_{B}-1) \Big] \\
	&+ 6 P_{B_j}^{3} \Bigg[ (1-A_2)\overline{S}_{B} \overline{\sigma}_{B}^3 +\Big[(1-3 A_2 + 2 A_1)\overline{N}_{B} -3(1-A_2) \Big]\overline{\sigma}_{B}^2 + \overline{N}_{B}(\overline{N}_{B}-1)\big[ (2A_1 -A_2)\overline{N}_{B} + 2A_2\big]
\Bigg]  \\
	  &+P_{B_j}^{4} \Bigg[(1-A_3)(K_B+3)\overline{\sigma}^4_{B} + \big[ 4(A_2 - A_3)\overline{N}_{B}-6(1-A_3)\big]S_{B}\overline{\sigma}_{B}^3 \\
   &\hspace{30pt} + \big[ (12A_2-6A_3-6A_1)\overline{N}^2_{B} + (18A_3-12A_2-6)\overline{N}_{B} + 11(1-A_3) \big] \overline{\sigma}^2_{B} \\
   &\hspace{30pt} + (4A_2-6A_1-A_3)\overline{N}_{B}^4 + (6A_1+6A_3-12A_2)\overline{N}_{B}^3 + (8A_2-11A_3)\overline{N}_{B}^2 + 6A_3 \overline{N}_{B}
\Bigg]
\Bigg\},
\end{split}
\end{equation}
\end{widetext}
where three coefficients $A_1$, $A_2$ and $A_3$ are 
\begin{equation}
	A_L=1- \prod_{k=1}^{L} \Bigg( \frac{\prod_{f} \prod_{i=1}^{n_{f,B_j}} \Big(1-k\frac{n_{f,B_j}}{N_f-i+1}\Big)}{\prod_{m=1}^{3} \Big( 1-k\frac{3}{N_q-m+1} \Big)}\Bigg)
	\label{A_coe}
\end{equation}
with $L=1,2,3$.

In the above formulas of variance, skewness and kurtosis, the first term in right hand side of the equation is always the dominant part and we find that it is just the result of binomial distribution with parameters ($\overline{N}_{B}$, $P_{B_j}$). In Fig.~\ref{fig2}, we plot $ \overline{\sigma}_{B_j}$, $\overline{S}_{B_j}$ and $\overline{K}_{B_j}$ of various identified baryons as the function of their production weights $P_{B_j}$. Symbols are full results and lines are binomial distributions as leading approximation. The size of quark system here is chosen to be $N_q=N_{\bar{q}}=500$ and the relative ratios of different quark flavors are set to be $N_{u}:N_{d}:N_{s}=1:1:0.43$. 
In addition, at large $\overline{N}_B$ and small $P_{B_i}$, binomial distribution converges toward the Poisson distribution. For multistrange hyperons such as $\Omega$ and $\Xi^{*}$,  their multiplicity distributions are well approximated by Poisson distribution because of quite small production weights $\sim 0.01$. However, multiplicity distributions of proton and $\Lambda$ can not be well approximated by Poisson distribution because of their relatively large production weights $\sim 0.1$. 

Multiplicity distribution of total baryons shows some slightly different properties from those of identified baryons. 
The variance of total baryon multiplicity is proportional to system size via $\overline{\sigma}^2_{B}/\overline{N}_{B} \approx 0.35$ at current baryon-meson competition and skewness is inversely proportional to system size via $\overline{S}_B \overline{N}^{1/2}_B \approx 0.37$. 
These properties are general expectations of stochastic combination process. 
But proportional coefficients can not to be explained in terms of the binomial distribution. 
This is easily understood. The number of quarks consumed by total baryon formation is about 20\% of total quark number in the system. This fact causes the deviation from the independent and stochastic feature of the binomial trial in each baryon production.

\subsection{two-baryon correlations}
\label{subsec_twoB}

Production of two different kinds of baryons is usually anti-associated in the hadronization of quark system with fixed quark numbers, characterized by the negative covariances of their multiplicities. The multiplicity covariance is defined as 
\begin{equation} 
	\overline{C}_{B_i B_j}=\overline{\delta N_{B_i} \delta N_{B_j}} =\overline{N_{B_i} N_{B_j}} - \overline{N_{B_i}}\, \overline{ N_{B_j} }.
\end{equation}
We consider two-baryon joint production
\begin{equation}
	\overline{N_{B_i}N_{B_j}}= P_{B_i B_j} \ \overline{N_{B}(N_{B}-1)}, 
	\label{nbibj}
\end{equation}
where the joint production probability of $B_i$$B_j$ pair can be evaluated by $ P_{B_i B_j}={N^{(q)}_{B_i B_j}}/{N_{6q}}$ with the number of six-quark cluster possible for $B_i$$B_j$ pair production $N^{(q)}_{B_i B_j} = N^{i}_{iter} N^{j}_{iter} \prod_{f} \prod_{i=1}^{ n_{f,B_i}+n_{f,B_j}} (N_f -i +1)$ and that for any two-baryon pair production $N_{6q}=\prod_{i=1}^{6}(N_q -i +1)$.

Substituting Eqs.~(\ref{avnbi}) and (\ref{nbibj}) into the covariance of two baryons, we get
\begin{equation}
	\frac{\overline{C}_{B_{i} B_{j}}}{\overline{N}_{B_{i}} \overline{N}_{B_{j}} } =\frac{P_{B_{i} B_{j}}}{P_{B_{i}} P_{B_{j}}} \frac{\overline{N_{B}(N_{B}-1)}}{\overline{N}_{B}^2}-1 ,
\end{equation}
in which
\begin{equation}
\begin{split}
	\frac{P_{B_{i} B_{j}}}{P_{B_{i}} P_{B_{j}}} &=\frac{ \prod_{f}\prod_{k=1}^{n_{f,B_{i}}} (1-\frac{n_{f,B_{j}}}{N_f -k +1}) }{\prod_{m=1}^{3} (1-\frac{3}{N_q -m +1}) } \\
	&= 1-  \sum_{f} n_{f,B_{i}} \, n_{f,B_{j}} \frac{1}{N_f} + \frac{9}{N_q} + \mathcal{O}(N^{-2}_q),
\end{split}
\end{equation}
where the product and summation of index $f$ run over all quark flavors and $n_{f,B_{i}}$ is the number of valance quark $f$ contained in hadron $B_{i}$.
Finally, we have 
\begin{equation}
	\frac{\overline{C}_{B_{i} B_{j}}}{\overline{N}_{B_{i}} \overline{N}_{B_{j}} } = - \sum_{f} \frac{n_{f,B_{i}} \, n_{f,B_{j}} }{N_f} - 
	\big( \frac{1}{\overline{N}_B} - \frac{\overline{\sigma}^2_{B}}{\overline{N}^{2}_{B}} - \frac{9}{N_q } \big) + \mathcal{O}(N_q^{-2}).
\label{cbba}
\end{equation}
The first part in the right hand side of the equation is the leading order contribution. It essentially originates from the fact that at hadronization once a quark enters into a $B_{i}$ it is consumed and therefore can not recombine into $B_{j}$. This part is inversely proportional to the quark number of the coincide flavor in two baryons, so the relative anti-correlation among strange baryons is usually greater than those of light flavor baryons. 
The part in bracket is the next-leading order contribution, which is usually a few percentages of the first part. It is negligible in correlations for the most baryon pairs with the coincide valance quark content but becomes important for correlations between baryon pairs with totally different quark flavors, such as $C_{p\Omega^{-}}$, $C_{\Delta^{++} \Delta^-}$, etc. 

\begin{figure}[!htbp]
  \centering
  \includegraphics[width=\linewidth]{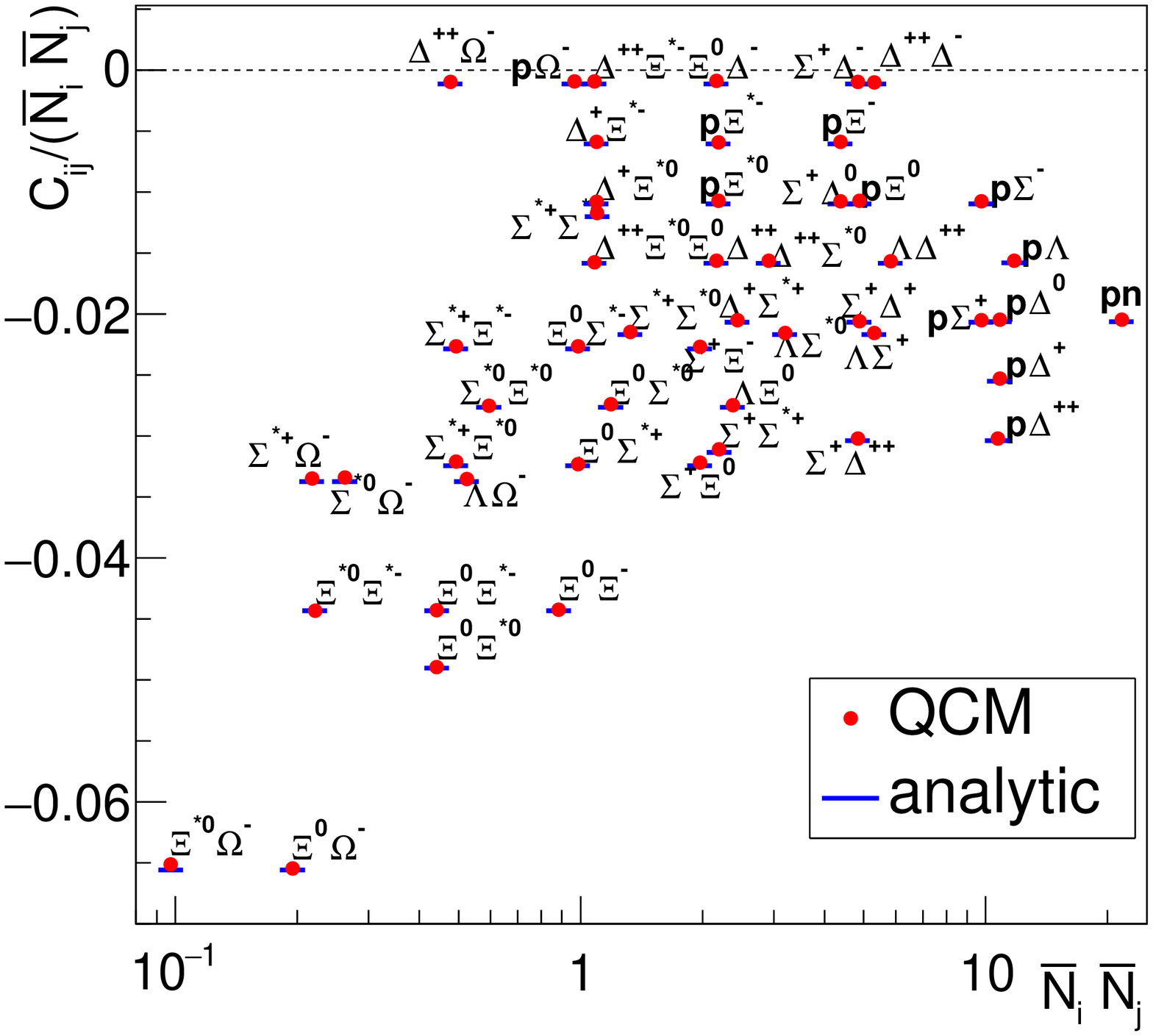}\\
  \caption{(Color online) Multiplicity covariance between two identified baryons. Different two-baryon pairs are distinguished in the horizontal axis by their multiplicity products. The size of quark system before hadronization is chosen to be $N_q = N_{\bar{q}} = 500$ and the relative ratios of different quark flavors are set to be $N_{u}:N_{d}:N_{s}=1:1:0.43$. Symbols are numerical results of QCM algorithm in Sec.~II and short solid lines are analytic results in Eq.~(\ref{cbba}).}
  \label{c2b_ini}
\end{figure}

In Fig.~\ref{c2b_ini}, we show results of the relative covariance $\overline{C}_{B_i B_j}/(\overline{N}_{B_{i}} \overline{N}_{B_{j}})$ of identified baryons produced from quark system hadronization with $N_q=N_{\bar{q}}=500$. Here, the relative ratios of different quark flavors are set to be $N_{u}:N_{d}:N_{s}=1:1:0.43$. 
Results of different two-baryon pairs are distinguished in the horizontal axis by their multiplicity products. Symbols are numerical results of QCM algorithm in Sec.~II and the short solid lines are analytic results in Eq.~(\ref{cbba}).
As discussed above, we see that the production of $\Xi\Omega^-$ and other hyperon pairs which share more strange content is the most anti-associated while those containing coincide light flavors are less anti-associated such as $pn$. 
For the $p\Omega^-$, $\Delta \Omega^-$, $\Xi^0 \Delta^-$ etc, there is no coincide flavor between two baryons but their productions are still anti-associated, although quite weak. This is due to the second term in right hand side of Eq.~(\ref{cbba}) and the physical origin is that the successive baryon production in combination process is suppressed by the baryon number conservation.

\subsection{baryon-antibaryon correlations}
\label{subsec_bbar}

It is generally expected that baryons and antibaryons are associated in their production, characterized by the positive covariance $\overline{C}_{B_i \bar{B}_{j}}=\overline{\delta N_{B_i} \delta N_{\bar{B}_j}} =\overline{N_{B_i} N_{\bar{B}_j}} - \overline{N_{B_i}}\, \overline{ N_{\bar{B}_j} }$ of their multiplicities. 
One main reason of this association comes from the global baryon number conservation in hadronization which is denoted by the quark number conservation 
$ N_B - N_{\bar{B}} = \frac{1}{3}\big(N_{q} - N_{\bar{q}}\big)$
in the combination process. This causes the following correlation between baryon and antibaryon
\begin{equation}
	\overline{C}_{B_i \bar{B}_{j} } = p_{B_i} p_{\bar{B}_{j}}\, \overline{\delta N_B  \, \delta N_{\bar{B}} } =  p_{B_i} p_{\bar{B}_{j}}\, \overline{\sigma}^2_{B},
\end{equation}
where we use $\overline{\delta N_B  \, \delta N_{\bar{B}} }=\overline{\sigma}^{2}_{B}=\overline{\sigma}^{2}_{\bar{B}}$ at fixed quark numbers. Inserting posterior production weight $p_{B_i}=\langle N_{B_i}\rangle / \langle N_{B} \rangle$, we get a scaling property 
\begin{equation}
	\frac{ \overline{C}_{B_i \bar{B}_{j} } }{ \overline{N}_{B_i} \overline{N}_{\bar{B}_j} } = \frac{ \overline{\sigma}^{2}_{B} }{ \overline{N}_{B} \overline{N}_{\bar{B}} }
	\label{cbbar}
\end{equation}
for baryon-antibaryon multiplicity correlations. 

\begin{figure}[!htbp]
	  \centering
	    \includegraphics[width=\linewidth]{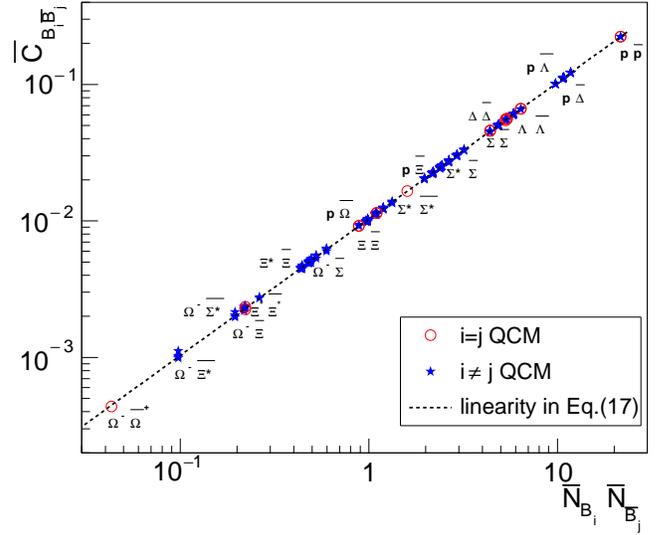}\\
		  \caption{(Color online) Multiplicity covariance between identified baryons and antibaryons. Different baryon-antibaryon pairs are distinguished in the horizontal axis by their multiplicity products. The size of quark system before hadronization is chosen to be $N_q = N_{\bar{q}} = 500$ and the relative ratios of different quark flavors are set to be $N_{u}:N_{d}:N_{s}=1:1:0.43$. Symbols are numerical results of QCM algorithm in Sec.~II and the dashed line is the scaling given by Eq.~(\ref{cbbar}).}
\label{cab_ini}
\end{figure}

In Fig.~\ref{cab_ini}, we compare the above formula with numerical results obtained from the algorithm in Sec.~II with quark system $N_q=N_{\bar{q}}=500$ in which relative ratios of different quark flavors are set to be $N_{u}:N_{d}:N_{s}=1:1:0.43$. The well agreement suggests that global baryon number conservation is the dominant reason for the production correlation between identified baryons and antibaryons.  
One interesting result is that both $\overline{C}_{B_i \bar{B}_{i}}$ and $\overline{C}_{B_i \bar{B}_{j}}$ ($i\neq j$) follow the same scaling line, which indicates that the production of baryon-antibyaron pair does not suffer more important constrain than that of two different baryons. This is reasonable in the case of free combination of quarks and antiquarks. Using the $\overline{N}_B/\overline{N}_M \approx 1/12$ and $N_s/N_u \approx 0.43$ which reproduce yield data in relativistic heavy ion collisions, we can estimate $\overline{N}_{B} \approx \frac{1}{6} N_u \approx \frac{1}{3} N_s$, which means the baryon number conservation is the strongest constraint in baryon-antibaryon joint production. 
The production of identified baryon $B_i$ and antibaryon $\bar{B}_j$ consumes only small fraction of total quarks and antiquarks and thus does not reach the conservation threshold of specific quark flavors.

\subsection{multi-body correlations}
Following the similar procedure, we also get multi-baryon correlations due to the exclusion effect of successive baryon production discussed in Sec.~\ref{subsec_twoB} and baryon number conservation in baryon/antibaryon production in Sec.~\ref{subsec_bbar}.  The three-baryon correlation is 
\begin{equation}
\begin{split}
	\overline{C}_{\alpha \beta \gamma } & = \overline{\delta N_{\alpha} \delta N_{\beta}\delta N_{\gamma}  }  \\
	& = \overline{N_{\alpha} N_{\beta}N_{\gamma}  } - \overline{N}_{\alpha} \overline{N}_{\beta} \overline{N}_{\gamma} - \overline{N}_{\alpha}\overline{C}_{\beta\gamma}-\overline{N}_{\beta}\overline{C}_{\alpha\gamma}-\overline{N}_{\gamma}\overline{C}_{\alpha\beta},
\end{split}
\end{equation}
and four-baryon correlation is 
\begin{equation}
\begin{split}
	\overline{C}_{\alpha \beta \gamma \epsilon } & = \overline{\delta N_{\alpha} \delta N_{\beta}\delta N_{\gamma} \delta N_{\epsilon}  }  \\
	& = \overline{N_{\alpha} N_{\beta}N_{\gamma} N_{\epsilon} } - \overline{N}_{\alpha} \overline{N}_{\beta} \overline{N}_{\gamma}  \overline{N}_{\epsilon}- \overline{N}_{\alpha}\overline{C}_{\beta\gamma\epsilon} -\overline{N}_{\beta}\overline{C}_{\alpha\gamma\epsilon} \\
	&- \overline{N}_{\gamma}\overline{C}_{\alpha\beta\epsilon} - \overline{N}_{\epsilon}\overline{C}_{\alpha\beta\gamma} -\overline{N}_{\alpha} \overline{N}_{\beta} \overline{C}_{\gamma\epsilon} -\overline{N}_{\alpha} \overline{N}_{\gamma} \overline{C}_{\beta\epsilon} \\
	&-\overline{N}_{\alpha} \overline{N}_{\epsilon} \overline{C}_{\beta\gamma} -\overline{N}_{\beta} \overline{N}_{\gamma} \overline{C}_{\alpha\epsilon} -\overline{N}_{\beta} \overline{N}_{\epsilon} \overline{C}_{\alpha\gamma} -\overline{N}_{\gamma} \overline{N}_{\epsilon} \overline{C}_{\alpha\beta}. 
\end{split}
\end{equation}

The average multiplicity product of three baryons ($\alpha\beta\gamma\in $ baryon) can be written as
\begin{equation}
\begin{split}
	\overline{N_{\alpha} N_{\beta}N_{\gamma}} &=  (1-A_{\alpha \beta \gamma} )\, \overline{N}_{\alpha} \overline{N}_{\beta} \overline{N}_{\gamma}  \\
	& \times \frac{ \overline{\delta N^3_{B}}+ 3 \overline{\sigma}^2_{B}(\overline{N}_{B} -1)+\overline{N}_{B}(\overline{N}_{B} -1)(\overline{N}_{B} -2)}{\overline{N}^3_B} \\
	& + \delta_{\alpha \beta} (1-\delta_{\alpha \gamma}) (\overline{C}_{\alpha \gamma} + \overline{N}_{\alpha} \overline{N}_{\gamma})  \\
	& +  (\delta_{\alpha \gamma} + \delta_{\beta \gamma}) (1-\delta_{\alpha \beta}) (\overline{C}_{\alpha \beta} + \overline{N}_{\alpha} \overline{N}_{\beta}) \\
	& + \delta_{\alpha \beta} \  \delta_{\alpha \gamma} \big[ 3\overline{\sigma}_{\alpha}^{2} + 3 \overline{N}_{\alpha} ( \overline{N}_{\alpha} -1 ) + \overline{N}_{\alpha} \big],
\end{split}
	\label{amp3b}
\end{equation}
where $\delta_{\alpha \beta}$ is Kronecker delta function and $\overline{\delta N^3_{B}}\equiv \overline{S}_B \overline{\sigma}^3_B $ is the third moment of total baryons. 
For that of two baryons and one antibaryon, e.g.,~$\alpha\beta\in \text{baryon}$, $\bar{\gamma} \in \text{antibaryon}$,  we have 
\begin{widetext}
\begin{equation}
\begin{split}
	\overline{N_{\alpha} N_{\beta}N_{\bar{\gamma}}} &=  (1-A_{\alpha \beta } )\, \overline{N}_{\alpha} \overline{N}_{\beta} \overline{N}_{\bar{\gamma}}  
	\times \frac{ \overline{\delta N^3_{B}}+  \overline{\sigma}^2_{B}(3\overline{N}_{B} +2c -1)+\overline{N}_{B}(\overline{N}_{B} -1)\overline{N}_{\bar{B}} }{\overline{N}^2_B \overline{N}_{\bar{B}}} 
	+ \delta_{\alpha \beta} ( \overline{C}_{\alpha \bar{\gamma}} + \overline{N}_{\alpha} \overline{N}_{\bar{\gamma}} ).
\end{split}
\end{equation}
Here, $c=\overline{N}_{B}-\overline{N}_{\bar{B}}$ is the number of net baryons and taken to be zero at LHC. 

The average multiplicity product of four baryons ($\alpha\beta\gamma \epsilon \in $ baryon) can be written as
\begin{equation}
\begin{split}
	\overline{N_{\alpha} N_{\beta}N_{\gamma} N_{\epsilon}} &=  (1-A_{\alpha \beta \gamma \epsilon} )\, \overline{N}_{\alpha} \overline{N}_{\beta} \overline{N}_{\gamma}  \overline{N}_{\epsilon}\, \frac{1}{\overline{N}^4_B}  \times \Big\{ \overline{\delta N^{4}_{B}} 
	 +  (4\overline{N}_{B} -6)\overline{\delta N^{3}_{B}} +   \overline{\sigma}^2_{B}(6\overline{N}^2_{B}- 18\overline{N}_{B} + 11) \\
	&+\overline{N}_{B}(\overline{N}_{B} -1)(\overline{N}_{B} -2) (\overline{N}_{B} -3) \Big\} 
	 + \delta_{\alpha\beta} \delta_{\alpha\gamma} \delta_{\alpha\epsilon} ( 6 \overline{N_{\alpha}^{3}} - 11 \overline{N_{\alpha}^{2}} + 6 \overline{N}_{\alpha} )
	 + \delta_{\alpha\beta} \delta_{\alpha\gamma} (1-\delta_{\alpha\epsilon}) ( 3 \overline{N_{\alpha}^{2} N_{\epsilon}} - 2 \overline{ N_{\alpha} N_{\epsilon}} )\\
	& + \delta_{\alpha\beta} \delta_{\alpha\epsilon} (1-\delta_{\alpha\gamma}) ( 3 \overline{N_{\alpha}^{2} N_{\gamma}} - 2 \overline{ N_{\alpha} N_{\gamma}} )
	 + (\delta_{\alpha\gamma} \delta_{\alpha\epsilon} + \delta_{\beta\gamma} \delta_{\beta\epsilon}) (1-\delta_{\alpha\beta}) ( 3 \overline{N_{\alpha}^{2} N_{\beta}} - 2 \overline{ N_{\alpha} N_{\beta}} )\\
	& + \delta_{\alpha\beta} \delta_{\gamma\epsilon} (1-\delta_{\alpha\gamma}) ( \overline{N_{\alpha}^{2} N_{\gamma}} + \overline{N_{\alpha} N_{\gamma}^{2}} - \overline{ N_{\alpha} N_{\gamma}} )
	 + (\delta_{\alpha\gamma} \delta_{\beta\epsilon} +\delta_{\alpha\epsilon} \delta_{\beta\gamma}) (1-\delta_{\alpha\beta}) ( \overline{N_{\alpha}^{2} N_{\beta}} + \overline{N_{\alpha} N_{\beta}^{2}} - \overline{ N_{\alpha} N_{\beta}} )\\
	& + \delta_{\alpha\beta} (1-\delta_{\alpha\gamma}) (1-\delta_{\alpha\epsilon})(1-\delta_{\gamma\epsilon}) \overline{N_{\alpha} N_{\gamma} N_{\epsilon}}
	 + (\delta_{\alpha\gamma}+\delta_{\beta\gamma}) (1-\delta_{\alpha\beta}) (1-\delta_{\alpha\epsilon})(1-\delta_{\beta\epsilon}) \overline{N_{\alpha} N_{\beta} N_{\epsilon}}\\
	& + (\delta_{\alpha\epsilon}+\delta_{\beta\epsilon}+ \delta_{\gamma\epsilon})  (1-\delta_{\alpha\beta}) (1-\delta_{\alpha\gamma})(1-\delta_{\beta\gamma}) \overline{N_{\alpha} N_{\beta} N_{\gamma}},
\end{split}
\end{equation}
	where $\overline{\delta N^{4}_{B}}\equiv (\overline{K}_B+3) \overline{\sigma}^4_B $ is the fourth moment of total baryons. The average multiplicity product of three or two baryons can be read from Eqs.~(\ref{amp3b}) and (\ref{nbibj}). 
For that of three baryons and one antibaryon, e.g.,~$\alpha\beta \gamma\in \text{baryon}$, $\bar{\epsilon} \in \text{antibaryon}$, we have 
\begin{equation}
\begin{split}
	\overline{N_{\alpha} N_{\beta}N_{\gamma} N_{\bar{\epsilon}}} &=  (1-A_{\alpha \beta \gamma } )\, \overline{N}_{\alpha} \overline{N}_{\beta} \overline{N}_{\gamma}  \overline{N}_{\bar{\epsilon}}\, \frac{1}{\overline{N}^3_B \overline{N}_{\bar{B}}}  \times \Big\{ \overline{\delta N^{4}_{B}} 
	 +  (4\overline{N}_{\bar{B}}+3c-3)\overline{\delta N^{3}_{B}}  
	 + \overline{\sigma}^2_{B}\bigg[6\overline{N}^2_{\bar{B}}+9(c-1)\overline{N}_{\bar{B}} + 3c^2-6c+2\bigg] \\
	&+\overline{N}_{B}(\overline{N}_{B} -1)(\overline{N}_{B} -2) \overline{N}_{\bar{B}}  \Big\}
	+ \delta_{\alpha\beta}\delta_{\alpha\gamma}(3\overline{N_{\alpha}^{2} N_{\bar{\epsilon}}} -2 \overline{N_{\alpha}N_{\bar{\epsilon}}})
	+ \delta_{\alpha\beta}(1-\delta_{\alpha\gamma})\overline{N_{\alpha}N_{\gamma}N_{\bar{\epsilon}}}
	+ (\delta_{\alpha\gamma}+ \delta_{\beta\gamma})(1-\delta_{\alpha\beta})\overline{N_{\alpha}N_{\beta}N_{\bar{\epsilon}}},
\end{split}
\end{equation}
and for that of two baryons and two antibaryons,  e.g.,~$\alpha\beta \in \text{baryon}$, $\bar{\gamma}\bar{\epsilon} \in \text{antibaryon}$, we have
\begin{equation}
\begin{split}
	\overline{N_{\alpha} N_{\beta}N_{\bar{\gamma}} N_{\bar{\epsilon}}} &=  (1-A_{\alpha \beta } ) (1-A_{\bar{\gamma}\bar{\epsilon}} ) \overline{N}_{\alpha} \overline{N}_{\beta} \overline{N}_{\bar{\gamma}}  \overline{N}_{\bar{\epsilon}}\, \frac{1}{\overline{N}^2_B \overline{N}^{2}_{\bar{B}}}  \times \Big\{ \overline{\delta N^{4}_{B}} +  (4\overline{N}_{\bar{B}}+2c-2)\overline{\delta N^{3}_{B}}  
	 + \overline{\sigma}^2_{B}\bigg[6\overline{N}^2_{\bar{B}}+6(c-1)\overline{N}_{\bar{B}} + c^2-3c+1\bigg] \\
	&+\overline{N}_{B}(\overline{N}_{B} -1) \overline{N}_{\bar{B}} (\overline{N}_{\bar{B}} -1) \Big\}
	 + \delta_{\alpha\beta}\delta_{\bar{\gamma}\bar{\epsilon}}(\overline{N_{\alpha}^{2}N_{\bar{\gamma}}} + \overline{N_{\alpha}N_{\bar{\gamma}}^{2}} + \overline{N_{\alpha}N_{\bar{\gamma}}}) 
	 + \delta_{\alpha\beta}(1-\delta_{\bar{\gamma}\bar{\epsilon}})\,\overline{N_{\alpha}N_{\bar{\gamma}}N_{\bar{\epsilon}}} + \delta_{\bar{\gamma}\bar{\epsilon}}\,(1- \delta_{\alpha\beta})\,\overline{N_{\alpha}N_{\beta}N_{\bar{\gamma}}}.
\end{split}
\end{equation}
\end{widetext}
Coefficients $A_{\alpha\beta}$, $A_{\alpha\beta\gamma}$ and $A_{\alpha\beta\gamma\epsilon}$ are extension of Eq.(\ref{A_coe}), 
\begin{equation}
	A_{\alpha\beta\gamma\epsilon} =1-  \frac{ \prod_{f}\prod_{h=\beta}^{\epsilon}  \prod_{i=1}^{n_{f,h}} \Big(1-\frac{\sum_{h'=\alpha}^{h-1} n_{f,h'}}{N_f-i+1}\Big)}{\prod_{k=1}^{n_h-1} \prod_{m=1}^{3} \Big( 1-k\frac{3}{N_q-m+1} \Big)}.
\end{equation}
Here $n_h$ in the denominator denotes the number of involved baryons, i.e.,~$n_h=4$ for $\alpha\beta\gamma\epsilon$ and 3 for $\alpha\beta\gamma$. $n_{f,h}$ is the number of valance quark of flavor $f$ contained in hadron $h$. $h-1$ in the numerator denotes the hadron before $h$ in combination $\alpha\beta\gamma\epsilon$. 
Taking the charge conjugation operation, we get coefficients of antibaryons.

We can check that the following normalization is satisfied, 
\begin{equation}
\begin{split}
	\sum_{\alpha\beta\gamma\in B} \overline{C}_{\alpha \beta \gamma }&=  \overline{\delta N^{3}_{B}}, \\
	\sum_{\alpha\beta\gamma\epsilon\in B} \overline{C}_{\alpha \beta \gamma \epsilon}& =  \overline{\delta N^{4}_{B}}, 
\end{split}
\end{equation}
and
\begin{equation}
\begin{split}
	\sum_{\alpha\beta\gamma\in B , \bar{B}} (-1)^m \overline{C}_{\alpha \beta \gamma }  &=  \overline{\delta (N_{B}-N_{\bar{B}})^3}=0, \\
	\sum_{\alpha\beta\gamma \epsilon \in B , \bar{B}} (-1)^m \overline{C}_{\alpha \beta \gamma \epsilon}  &=  \overline{\delta (N_{B}-N_{\bar{B}})^4}=0,  
\end{split}
\end{equation}
where $m$ denotes the number of antibaryons in $\alpha\beta\gamma$ and $\alpha\beta\gamma\epsilon$ combinations. 

\section{baryon production from the quark system with variational quark numbers}

In this section, we take into account effects of fluctuations and correlations of quark numbers in system before hadronization on multiple production of baryons and antibaryons. We firstly give the general procedure of including quark number fluctuations and correlations in hadronic observables and then show the specific formulas for moments and two-body correlations of baryons and antibaryons. Then we discuss properties of quark number fluctuations and correlations in the context of ultra-relativistic heavy ion collisions and we show numerical results of baryon moments, two-baryon correlations and baryon-antibaryon correlations.

\subsection{general formulas of including variational quark numbers }
The produced quark system in heavy ion collisions at a specific collision energy is always varied in size event-by-event and the number of quarks and that of antiquarks in system at hadronization should follow a certain distribution $\mathcal{P}(\{ N_{q_i}, N_{\bar{q}_i}\}; \{ \langle N_{q_i}\rangle,\langle N_{\bar{q}_i}\rangle \})$ around the event-average quark numbers $\langle N_{q_i} \rangle$ and antiquark numbers $\langle N_{\bar{q}_i} \rangle$,  where $q_i=u$, $d$, $s$ are considered in this paper. In QCM, the distribution includes also the possible contribution of small-amount dynamical production of newborn quarks and antiquarks during hadronization process due to the requirement of exact energy conservation and entropy increase \cite{sj10s}. 
The event average of a hadronic physical quantity $A_h$ is 
\begin{widetext}
\begin{equation}
\begin{split}
	\langle A_h \rangle &=  \sum_{\{ N_{h_j} \}} A_h \mathcal{P}(\{ N_{h_j}\};\{ \langle N_{q_i}\rangle,\langle N_{\bar{q}_i}\rangle \}) =  \sum_{\{ N_{q_i}, N_{\bar{q}_i}\}} \sum_{\{ N_{h_j} \}} A_h \ \mathcal{P}(\{ N_{h_j}\};\{ N_{q_i}, N_{\bar{q}_i}\}) \mathcal{P}(\{ N_{q_i}, N_{\bar{q}_i}\}; \{ \langle N_{q_i}\rangle,\langle N_{\bar{q}_i}\rangle \}) \\
		&=\sum_{\{ N_{q_i}, N_{\bar{q}_i}\}} \overline{A}_h  \ 
\mathcal{P}(\{ N_{q_i}, N_{\bar{q}_i}\}; \{ \langle N_{q_i}\rangle,\langle N_{\bar{q}_i}\rangle \}).
\end{split}
\label{Ah_eq0}
\end{equation}
If $\overline{A}_h$ is known already, we can expand it as Taylor series at the event average of quark numbers $\{ \langle N_{q_i}\rangle,\langle N_{\bar{q}_i}\rangle \}$
\begin{equation}
\overline{A}_h =  \left. \overline{A}_h\right|_{\substack{\langle N_{q_i}\rangle \\ \langle N_{\bar{q}_i}\rangle}} + \sum_{f_1} \left. \frac{\partial \overline{A}_h} {\partial N_{f_1}} \right|_{\substack{\langle N_{q_i}\rangle \\ \langle N_{\bar{q}_i}\rangle}} \delta N_{f_1}  +  \frac{1}{2}\sum_{f_1,f_2} \left. \frac{\partial^2 \overline{A}_h} {\partial N_{f_1} \partial N_{f_2}} \right|_{\substack{\langle N_{q_i}\rangle \\ \langle N_{\bar{q}_i}\rangle}} \delta N_{f_1} \delta N_{f_2}  + \frac{1}{3!}\sum_{f_1,f_2,f_3} \left. \frac{\partial^3 \overline{A}_h} {\partial N_{f_1} \partial N_{f_2} \partial N_{f_3}} \right|_{\substack{\langle N_{q_i}\rangle \\ \langle N_{\bar{q}_i}\rangle}} \delta N_{f_1} \delta N_{f_2} \delta N_{f_3} + \mathcal{O}(\delta^4),
\end{equation}
where indexes $f_1$, $f_2$ and $f_3$ run over all quark and antiquark flavors and $\delta N_{f_1} = N_{f_1} - \langle N_{f_1} \rangle$. The subscript $\langle N_{q_i}\rangle,\langle N_{\bar{q}_i}\rangle $ denotes the evaluation at event average point.
Substituting it into Eq.~(\ref{Ah_eq0}), we get 
\begin{equation}
\langle A_h \rangle =  \left. \overline{A}_h\right|_{\substack{\langle N_{q_i}\rangle \\ \langle N_{\bar{q}_i}\rangle}} +  \frac{1}{2}\sum_{f_1,f_2} \left. \frac{\partial^2 \overline{A}_h} {\partial N_{f_1} \partial N_{f_2}} \right|_{\substack{\langle N_{q_i}\rangle \\ \langle N_{\bar{q}_i}\rangle}} C_{f_1 f_2} + \frac{1}{3!}\sum_{f_1,f_2,f_3} \left. \frac{\partial^3 \overline{A}_h} {\partial N_{f_1} \partial N_{f_2} \partial N_{f_3}} \right|_{\substack{\langle N_{q_i}\rangle \\ \langle N_{\bar{q}_i}\rangle}} C_{f_1 f_2 f_3} + \mathcal{O}(\delta^4),
\label{enav}
\end{equation}
\end{widetext}
where $C_{f_1 f_2}= \langle \delta N_{f_1} \delta N_{f_2} \rangle$ and $C_{f_1 f_2 f_3} = \langle \delta N_{f_1} \delta N_{f_2} \delta N_{f_3}\rangle$ are two-body and three-body correlation functions of quarks and antiquarks, respectively.  
Then the influence of quark number distribution on hadronic quantities can be taken into account by the mean, two-body and multi-body correlations of quark numbers. 
In the following equations we drop the subscript $\langle N_{q_i}\rangle,\langle N_{\bar{q}_i}\rangle $ for convenience.

\subsection{formulas of identified baryons}

Using Eq.~(\ref{enav}), we first get the event average of baryon multiplicity 
\begin{equation}
	\langle N_{B_i} \rangle = \overline{N}_{B_i} +  \frac{1}{2}\sum_{f_1,f_2} \frac{\partial^2 \overline{N}_{B_i}} {\partial N_{f_1} \partial N_{f_2}} C_{f_1 f_2} + \mathcal{O}(N_f^{-2}).
	\label{nbi_qnfc}
\end{equation}
The effect of two-quark correlations on baryon multiplicity is the order of magnitude of $1/\langle N_f\rangle$, which is only a few percentages of the leading term due to the large quark number (i.e.,~hundreds of quarks and antiquarks per unit rapidity at RHIC and LHC energies). The influence of three-body and four-body correlations of quarks and antiquarks is suppressed further by  $1/N^2_f$. Therefore, effects of quark number correlations and fluctuations can be safely neglected in studies of inclusive multiplicities of identified hadrons in relativistic heavy ion collisions, as we did in previous works.

For moments of multiplicity distributions of identified hadrons, we have
\begin{equation}
	\sigma^2_{B_i} = \overline{\sigma}^2_{B_i} + \sum_{f_1,f_2} \bigg(  \partial_1  \overline{N}_{B_i}  \partial_2  \overline{N}_{B_i} +\frac{1}{2} \partial_{12}   \overline{\sigma}^2_{B_i}   \bigg)\, C_{f_1,f_2}+ \mathcal{O}(N_f^{-2}),
	\label{vbi_qnfc}
\end{equation}

\begin{widetext}
\begin{equation}
	S_{B_i}  = \overline{S}_{B_i} \  \bigg\{ 1  +  \sum_{f_1,f_2} \Big[ 
 \frac{ \partial_{12}\overline{\delta N^3_{B_i}} + 3 \partial_1 \overline{N}_{B_i} \partial_2 \overline{\sigma}^2_{B_i} +3 \partial_2 \overline{N}_{B_i} \partial_1 \overline{\sigma}^2_{B_i}}{2 \overline{\delta N^3_{B_i}}} 
		- 3 \frac{ (\partial_1 \overline{N}_{B_i})( \partial_2 \overline{N}_{B_i}) + \frac{1}{2} \partial_{12} \overline{\sigma}^2_{B_i} }{2 \overline{\sigma}^2_{B_i}} 
\Big] C_{f_1 f_2} + \mathcal{O}(N_f^{-2})
\bigg\}, 
\end{equation}
\begin{equation}
\begin{split}
	K_{B_i} = \overline{K}_{B_i} + (\overline{K}_{B_i} +3) \Bigg\{&  \sum_{f_1,f_2} \Big( \frac{ \partial_{12} \overline{\delta N^4_{B_i}} + 8  \partial_{1} \overline{\delta N^3_{B_i}}\partial_{2}\overline{N}_{B_i} +  12 \overline{\sigma}^2_{B_i} \partial_{1}\overline{N}_{B_i} \partial_{2}\overline{N}_{B_i} }{2  \overline{\delta N^4_{B_i}}} - 2 \frac{\partial_{1}\overline{N}_{B_i} \partial_{2}\overline{N}_{B_i} + \frac{1}{2} \partial_{12} \overline{\sigma}^2_{B_i} }{ \overline{\sigma}^2_{B_i} } \Big)C_{f_1 f_2}  + \mathcal{O}(N_f^{-2}) 
\Bigg\}.
\label{kbi_qnfc}
\end{split}
\end{equation}
\end{widetext}
Here, we have used $\partial_{1} \equiv \frac{\partial}{\partial N_{f_1}}$ and $\partial_{12} \equiv \frac{\partial^2}{\partial N_{f_1} \partial N_{f_2}}$ for abbreviation. 
Because higher order contributions of quark correlations and fluctuations are usually suppressed by the factor $1/\langle N_f\rangle$, here we only show effects of second order correlations and fluctuations of quark numbers on the directly produced baryons.

For two-body correlations of baryons and antibaryons, we have
\begin{equation}
	C_{\alpha \beta} = \overline{C}_{\alpha \beta} + \frac{1}{2} \sum_{f_1,f_2} \Big[  2 \partial_{1}\overline{N}_{\alpha} \partial_{2}\overline{N}_{\beta} +\partial_{12}  \overline{C}_{\alpha \beta} \Big] C_{f_1 f_2} + \mathcal{O}(N_f^{-2}).
	\label{c2_qnfc}
\end{equation}
Here, the contribution of second order quark correlations is the same order as $\overline{C}_{\alpha \beta}$, and they might cancel with each other significantly. The influence of higher order contributions of quark correlations is about few percentages at LHC and is neglected here.  As $\alpha=\beta$, we obtain Eq.~(\ref{vbi_qnfc}) which is also hardly influenced by higher order quark correlations. 

In appendix, we supplement the procedure of obtaining the full expression of Eqs.~(\ref{vbi_qnfc})-(\ref{c2_qnfc}) up to the four-body quark correlations for readers' convenience and decay calculations in the next section. 

\subsection{quark number correlations and fluctuations}

We firstly determine the size of quark system before hadronization which is consistent with that produced in relativistic heavy ion collisions at LHC energy. By fitting the rapidity density of hadronic yield in central Pb+Pb collisions at $\sqrt{s_{NN}}=$ 2.76 TeV, we obtain $\langle N_q \rangle = \langle N_{\bar{q}} \rangle =1710$ and the strangeness content $\langle N_s \rangle/\langle N_u \rangle= \langle N_s \rangle/\langle N_d \rangle = 0.43$ for quark system in unit rapidity window $y_w=1$ in central rapidity region. We note that the obtained strangeness suppression factor $\lambda_s \equiv \langle N_s \rangle/\langle N_u \rangle= \langle N_s \rangle/\langle N_d \rangle = 0.43$ is in agreement with the Wroblewski parameter calculated by Lattice QCD \cite{Gavai05,Mukherjee06}.  In the following sections, we use it as the default size of quark system. 
If different $y_w$ is selected, quark numbers in system are multiplied by factor $y_w$ because we always focus on the central rapidity plateau region $y_w<1.5$ where the rapidity distribution of quark numbers is uniform. 

For two-body correlation $C_{f_1 f_2}$ of quarks and antiquark, using the charge conjugation symmetry and isospin symmetry between $u$ and $d$ quarks for the quark system produced at LHC, there are only 8 relevant quark correlations, i.e.,
\begin{itemize}
	\item[] two variances $C_{uu} \equiv \sigma^2_u$ and $C_{ss} \equiv \sigma^2_s$,
	\item[] two pair correlations $C_{u\bar{u}}$ and $C_{s\bar{s}}$,
	\item[]	four off-diagonal correlations $C_{ud}$, $C_{us}$, $C_{u\bar{d}}$ and $C_{u\bar{s}}$. 
\end{itemize}
Variances of quark numbers are usually approximated to follow Poisson statistics $\sigma^2_u \approx \langle N_u\rangle$ and $\sigma^2_s \approx \langle N_s \rangle$ for a thermalized quark system with grand canonical ensemble. Lattice QCD calculations at vanishing chemical potential provide important constraint on the above quark correlations \cite{Dht15}, which show the weak off-diagonal flavor susceptibilities of quark numbers $\chi_{us}/\chi_{ss} \approx -0.05$  and $\chi_{ud}/\chi_{uu} \approx -0.05$ as temperature closes to the confinement phase boundary. Here, $\chi_{us} \equiv C_{us} + C_{\bar{u}\bar{s}} - C_{u\bar{s}} - C_{\bar{u}s} = 2(C_{us} - C_{u\bar{s}})$ and others are similarly defined.  
Because of the lack of further theoretical constraints on those quark number correlations at present, we have to adopt some symmetry approximations on quark correlations, i.e., $C_{u\bar{s}}/C_{us}=C_{u\bar{d}}/C_{ud}=\lambda_1$ and $C_{u\bar{u}}/\sigma^2_{u}=C_{s\bar{s}}/\sigma^2_s=\lambda_2$ where $\lambda_1$ and $\lambda_2$ are treated as parameters of this work. The value of $\lambda_2$ is smaller than one if we consider a slice of quark system, e.g.,~ mid-rapidity region, produced in heavy ion collisions. The off-diagonal flavor correlations are usually expected to be much smaller than variances of quark numbers. Inspired by the weak off-diagonal flavor susceptibilities in Lattice QCD calculations, we assume $C_{ud}/C_{uu} \sim 0.05$ (correspondingly $\lambda_1 \sim 2.0$) with some arbitrariness in this work to study effects of the weak flavor off-diagonal quark correlations on baryon and antibaryon production. 

Since this work focuses on the baryon sector, we introduce the total baryon number balance coefficient $\rho^{(q)}_B$ as one physical characteristic of the quark system,
\begin{equation}
\begin{split}
	\rho^{(q)}_B = \frac { \sum_{f_1,f_2} \frac{1}{3} C_{f_1 \bar{f}_2} }{ N^{(q)}_{B} } 
 = \lambda_2 -0.1 \lambda_1 \frac{1-\lambda_2}{1-\lambda_1} \frac{1+2\lambda_s}{2+\lambda_s},
\end{split}
\end{equation}
where indexes $f_1,f_2$ run over all flavors of quarks and $N^{(q)}_B= \frac{1}{3}(\langle N_u \rangle + \langle N_d \rangle + \langle N_s \rangle)$. Note that the factor ${1}/{3}$ before $C_{f_1\bar{f}_2}$ denotes the balanced baryon number if $f_1$ and $\bar{f}_2$ are correlated. The second equal uses the above approximated quark correlations. We also introduce the electric charge balance coefficient of quark system, which is defined as 
\begin{equation}
	\rho^{(q)}_C = \frac{1}{N^{(q)}_{C}} \sum_{f_1,f_2} min(Q_{f_1},Q_{f_2})\, C_{f_1 \bar{f}_2}, 
	\label{rho_c}
\end{equation}
where indexes $f_1,f_2 = u, \bar{d}, \bar{s}$ run over all positively charged quarks with electric charges $Q_{f_1}$ and $Q_{f_2}$, respectively. $N^{(q)}_C= \frac{1}{3}(2\langle N_u \rangle + \langle N_{\bar{d}} \rangle + \langle N_{\bar{s}} \rangle)$. The balanced charge for $f_1 \bar{f}_2$ pair is the minimum of their electric charges. Above approximated quark two-body correlations guarantee the correct boundary behavior of conserved charge for quark system, i.e.,~as $\lambda_2$ goes to one both $\rho^{(q)}_B$ and $\rho^{(q)}_C$ go to one.  Using the measured charge balance function of thermal particles in central Pb+Pb collisions at $\sqrt{s_{NN}}=2.76$ TeV \cite{AliceBF13}, we can roughly constrain the $\rho^{(q)}_C$ of quark system 
\begin{equation}
	\rho^{(q)}_C (y_w) \approx \int_{0}^{y_w} B(\delta \eta) d\delta \eta,
	\label{rho_c_bf}
\end{equation}
if we expect the small change of charge balance property of system during hadronization \cite{SJ12FB}. 
By $\rho^{(q)}_C (y_w)$ we can fix $\lambda_2$ and other off-diagonal elements of two-body correlations which are also dependent on $y_w$. 

\begin{figure*}[!thbp]
	\centering
		\includegraphics[width=\linewidth]{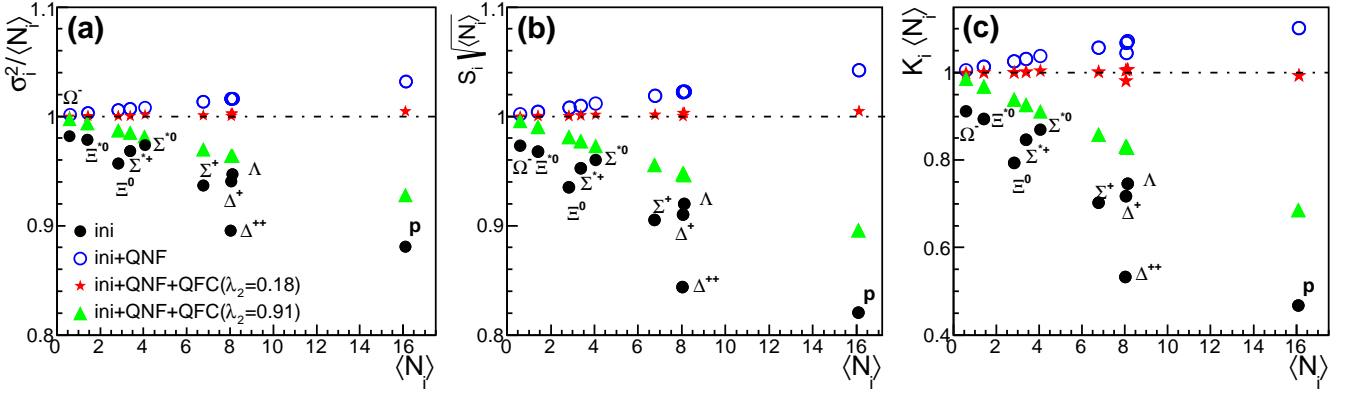}
		\caption{(Color online) Moments of identified baryons after considering quark number fluctuations (QNF) and quark number flavor conservation (QFC) with parameter $\lambda_2$. The $\lambda_2=0.91$ is chosen to be consistent with the observed charge balance function of thermal particles in unit pseudo-rapidity window observed in Pb+Pb collisions at 2.76 TeV \cite{AliceBF13}.}
		\label{bmoment2}
\end{figure*}

Three-body and four-body correlations of quarks and antiquarks influence relatively less on the physical quantities of initial baryons in previous subsection than the two-body correlations of quark numbers. But they will influence those of final baryons through resonance decays (as shown in next section ), so we need them also. Because there are no theoretical calculations at present which we can borrow, we take the following approximation for three-body quark correlations, i.e.,~$C_{fff} \equiv \langle \delta N_f^3 \rangle =\langle N_f \rangle $ and off-diagonal correlations $C_{f_{1}f_{2}f_{3}}=0$ where $f_{1}$, $f_{2}$ and $f_{3}$ are different flavors. For four-body correlations, we approximate them using two-body correlations
\begin{equation}
\begin{split}
	C_{f_{1}f_{2}f_{3}f_{4}} \approx  &C_{f_{1}f_{2}} C_{f_{3}f_{4}} + C_{f_{1}f_{3}} C_{f_{2}f_{4}}+C_{f_{1}f_{4}} C_{f_{2}f_{3}} \\
	&{} + 3 \delta_{f_{1}f_{3}}\delta_{f_{2}f_{4}} C_{f_{1}f_{2}} \\
	&{} + 3 \delta_{f_{1}f_{4}}\delta_{f_{2}f_{3}} C_{f_{1}f_{2}} \\
	&{} + 3 \delta_{f_{1}f_{3}}\delta_{f_{2}f_{4}} C_{f_{1}f_{3}}, 
\end{split}
\end{equation}
\begin{equation}
\begin{split}
	C_{\bar{f_{1}}f_{2}f_{3}f_{4}} \approx  &C_{\bar{f_{1}}f_{2}} C_{f_{3}f_{4}} + C_{\bar{f_{1}}f_{3}} C_{f_{2}f_{4}}+C_{\bar{f_{1}}f_{4}} C_{f_{2}f_{3}} \\
	&+ 9\delta_{f_{2}f_{3}}\delta_{f_{2}f_{4}}C_{\bar{f_{1}}f_{2}}, 
\end{split}
\end{equation}
\begin{equation}
\begin{split}
	C_{\bar{f_{1}}\bar{f_{2}}f_{3}f_{4}} \approx  &C_{\bar{f_{1}}\bar{f_{2}}} C_{f_{3}f_{4}} + C_{\bar{f_{1}}f_{3}} C_{\bar{f_{2}}f_{4}}+C_{\bar{f_{1}}f_{4}} C_{\bar{f_{2}}f_{3}} \\
	&+ 9\delta_{\bar{f_{1}}\bar{f_{2}}}\delta_{f_{3}f_{4}}C_{\bar{f_{1}}f_{3}}. 
\end{split}
\end{equation}
By this approximation, the kurtosis of net baryons has the property $K_{netB}~\sigma^2_{netB}=1$ which is suggested in ultra-relativistic heavy ion collisions \cite{star12Moments}. 

\subsection{numerical results of multiplicity moments of identified baryons}

Fig.~\ref{bmoment2} shows moments of various identified baryons after taking into account effects of quark number correlations and fluctuations.
The system size is taken to be the default value of unit $y_w$.
In order to clearly present effects of quark correlations and fluctuations, the variance $\sigma^2_{i}$, skewness $S_i$, and kurtosis $K_i$ of identified baryons are multiplied by factors $1/\langle N_i \rangle$, $\sqrt{\langle N_i \rangle}$, and $\langle N_i \rangle$, respectively, to make them the order of one. Here, the usage of $\langle N_{i} \rangle$ as the scaling factor is due to its insensitivity to correlations and fluctuations of quark numbers. 
We present results caused by quark combination process (marked by ``ini''), results including effects of quark number fluctuations (marked by ``ini+QNF''), and results further including effects of quark flavor conservation (marked by ``ini+QNF+QFC''). The last case is the physical result.  The purpose of such presentation is to show the contributions of different sources in final physical results.

Solid circles in Fig.~\ref{bmoment2}(a) are the variance of initial baryons directly produced by hadronization. As discussed previously, $\sigma^2_{i}/\langle N_i \rangle$ of identified baryons, roughly following binomial distribution, is always smaller than one and usually decreases with the increase of multiplicity or production weight. $\Omega^{-}$ is only 2\% smaller than one while proton about 10\%. But there are several exceptions for such a decreasing trend. For example, variance of $\Delta^{++}$ is smaller than its isospin partner $\Delta^{+}$ although their multiplicities are nearly same. This is due to the effect of identical quark flavor in baryon production encoded via coefficient $A_L$ in their variance formula in Eq.~(\ref{iniSigBi}). Others exceptions including those between $\Xi$ and $\Sigma^{*}$ and those between $\Sigma^{+}$ and $\Lambda$ are due to the same reasons either in strange or light flavor sector. 
These properties are also observed in baryon's skewness Fig.~\ref{bmoment2}(b) and kurtosis Fig.~\ref{bmoment2}(c) with larger amplitude.

Open circles show the baryon moments after considering effects of quark number fluctuations. We can see that fluctuations of quark numbers obviously increase the baryon's multiplicity fluctuations. $\sigma^2_{i}/\langle N_i \rangle$ of various baryons exceeds one. Proton is about 3\% greater than one while $\Omega^{-}$ also slightly exceeds one. Skewness and kurtosis of baryons are also greater than one and they are more sensitive to quark number fluctuations, e.g.,~proton skewness increases about 5\% and kurtosis about 10\%, respectively. 
The numerical reason of such rapid increase, taking variance for example, is that quark number fluctuations contribute to baryon variance mainly via $\sum_f ({\partial \overline{N}_{B_i}}/{\partial N_f} )^2 \sigma^2_f$ term in Eq.~(\ref{vbi_qnfc}) but contribute to baryon yield via $\sum_f ({\partial^2 \overline{N}_{B_i}}/{\partial N^2_f} )  \sigma^2_f$ term in Eq.~(\ref{nbi_qnfc}) which is much smaller than the former. We emphasize that these results are not the final physical predictions of baryon moments because we should always consider the effect of flavor (or charge) conservation in the studied rapidity window in the context of relativistic heavy ion collisions. 

Solid up-triangles show baryon moments after considering further effects of quark flavor conservation with parameter $\lambda_2 =0.91$, besides of quark number fluctuations. Here the value of parameter $\lambda_2$  is chosen so that the electric charge balance coefficient $\rho^{(q)}_C$of quark system, according to Eq.~(\ref{rho_c_bf}), is consistent with the measured charge balance function in unit pseudo-rapidity window in central Pb+Pb collisions at $\sqrt{s_{NN}}=2.76$ GeV \cite{AliceBF13}.  
Considering the pair association of quark and antiquark will facilitate meson production and suppress baryon production. 
Comparing to open circles, we therefore observe a significant decrease of proton variance, skewness and kurtosis. 
Such influence of flavor (or charge) conservation has been studied in Ref. \cite{koch13}.
For baryons with small multiplicities such as $\Omega^{-}$ and $\Xi^{*}$, they are weakly influenced by flavor conservation of quark numbers and their moments are always almost one.
If we choose smaller flavor conservation parameter $\lambda_2 = 0.18$ which corresponds to the observed charge balance in small rapidity window $y_w \approx 0.15$, we can observe almost unitary baryon's moments, shown as star symbols,  which is similar to Poisson distribution. 
However, for such small $y_w$, particle exchange in the window boundary due to the rapidity shift in hadronization, resonance decays and particle rescatterings is significant and therefore statistic effect is dominant. Poisson distribution is then usually expected but the microscopic dynamics of hadron production is lost at such small $y_w$.

\subsection{numerical results of two-baryon correlations}

\begin{figure}[!htbp]
\centering
  \includegraphics[width=\linewidth]{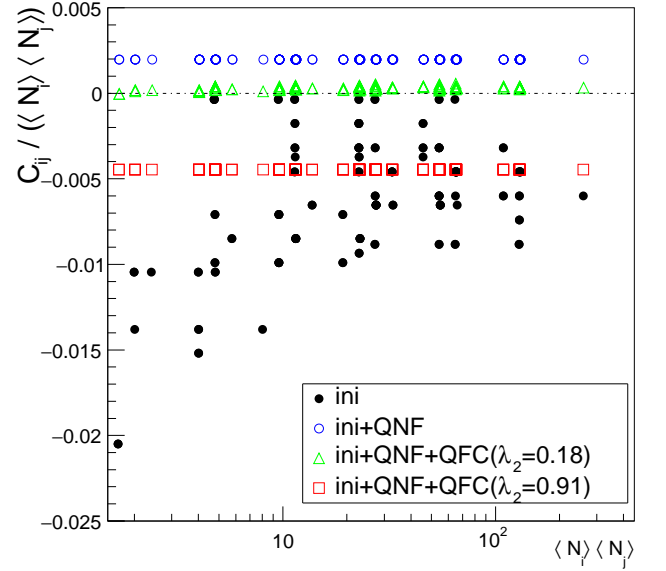}\\
	\caption{(Color online) Two-baryon multiplicity correlations after considering the effects of quark number fluctuations (QNF) and quark flavor conservation (QFC) just before hadronization. The labels for solid circles are the same as those in Fig.~\ref{c2b_ini}.  }
  \label{c2b_qnfc}
\end{figure}

Fig.~\ref{c2b_qnfc} shows two-baryon multiplicity correlations after considering effects of quark number fluctuations and correlations. 
The system size is taken to be the default value of unit $y_w$.
Solid circles show initial two-baryon correlations due to the hadronization of the quark system with given quark numbers and antiquark numbers. They exhibit a sensitive dependence on baryon species, as discussed in detail in Sec.~\ref{subsec_twoB}. After taking into account effects of quark number fluctuations, all two-baryon correlations, open circles, flip the sign and become a positive and almost universal value. The positive value means the production of two baryons is associated, which is because that both two baryons parallelly respond to the change of quark numbers or that of antiquark numbers. This association is suppressed and/or canceled by further taking into account the flavor conservation of quark numbers. 
With small flavor conservation parameter $\lambda_2=0.18$, all two-baryon correlations, open up-triangles, tend to be zero. With practical $\lambda_2 =0.91$ for unit rapidity window size, we get the physical prediction of two-baryon correlations shown as open squares. We see a strong production anti-association between two baryons, and interestingly we see a universal value for all two-baryon correlations. This is a striking characteristic of two-baryon production in QCM.

\begin{figure}[!htbp]
\centering
  \includegraphics[width=\linewidth]{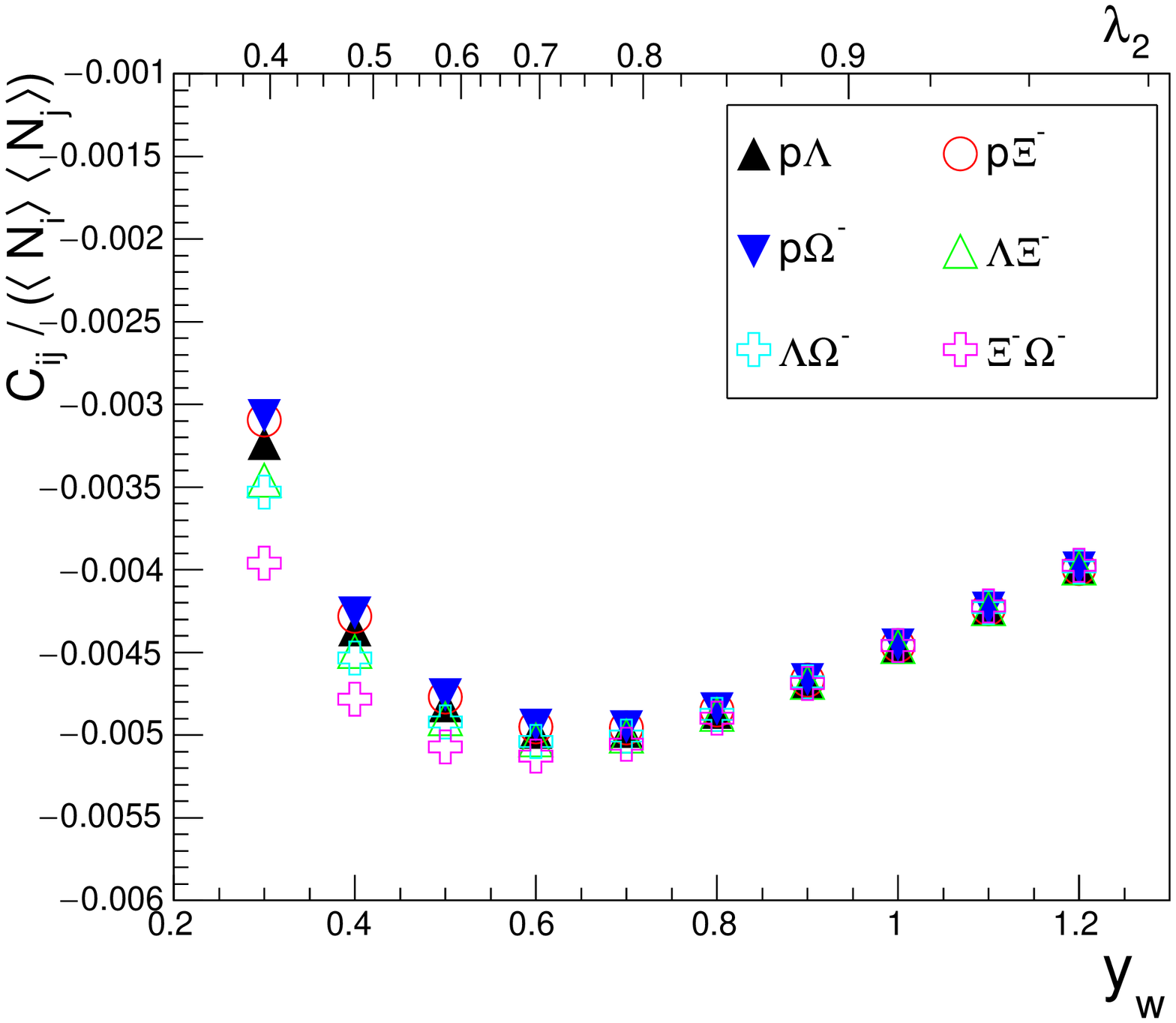}\\
  \caption{(Color online) Rapidity window size dependence of two-baryon correlations after taking into account effects of quark number correlations and fluctuations before hadronization. The auxiliary horizontal axis on top of figure shows the corresponding value of flavor conservation parameter $\lambda_2$. }
  \label{c2b_stable_qnfc}
\end{figure}

Fig.~\ref{c2b_stable_qnfc} shows two-body correlations of stable baryons $p$, $\Lambda$, $\Xi^{-}$, $\Omega^{-}$ at different rapidity window sizes $y_w$. 
In order to closely relate to the experimental measurement at specific $y_w$, we have introduced the electric charge balance coefficient of quark system $\rho^{(q)}_c$ defined in Eq.~(\ref{rho_c}), and we estimate its value by Eq.~(\ref{rho_c_bf}) using the data of charge balance function \cite{AliceBF13}. After obtaining the $\rho^{(q)}_c(y_w)$, we fix the flavor  conservation parameter $\lambda_2(y_w)$.  The value of $\lambda_2$ as the function of $y_w$ is shown as auxiliary horizontal axis on top of figure. 
Note that the average quark numbers of quark system are also linearly changed with $y_w$. 
We see a nonmonotonic behavior of two-baryon correlations with respect to $y_w$, which is due to the competition between the changed flavor conservation and the changed quark numbers of system. 
As $y_w$ increases from 0.3 to 0.6, the flavor conservation coefficient $\lambda_2$ increases rapidly up to about 0.7 and this leads to the increased anti-association between two baryons.
However, as $y_w$ continues to enlarge, the effect of increased flavor conservation is overwhelmed by that of the increased quark numbers and we see a decreased anti-association between two baryons. We also see that with the increased $y_w$ the difference between different two-baryon correlations decreases and we have an almost universal correlation magnitude for all two-baryon correlations, as shown in Fig.~\ref{c2b_qnfc}.

\subsection{numerical results of baryon-antibaryon correlations}

\begin{figure}[!htbp]
\centering
  \includegraphics[width=\linewidth]{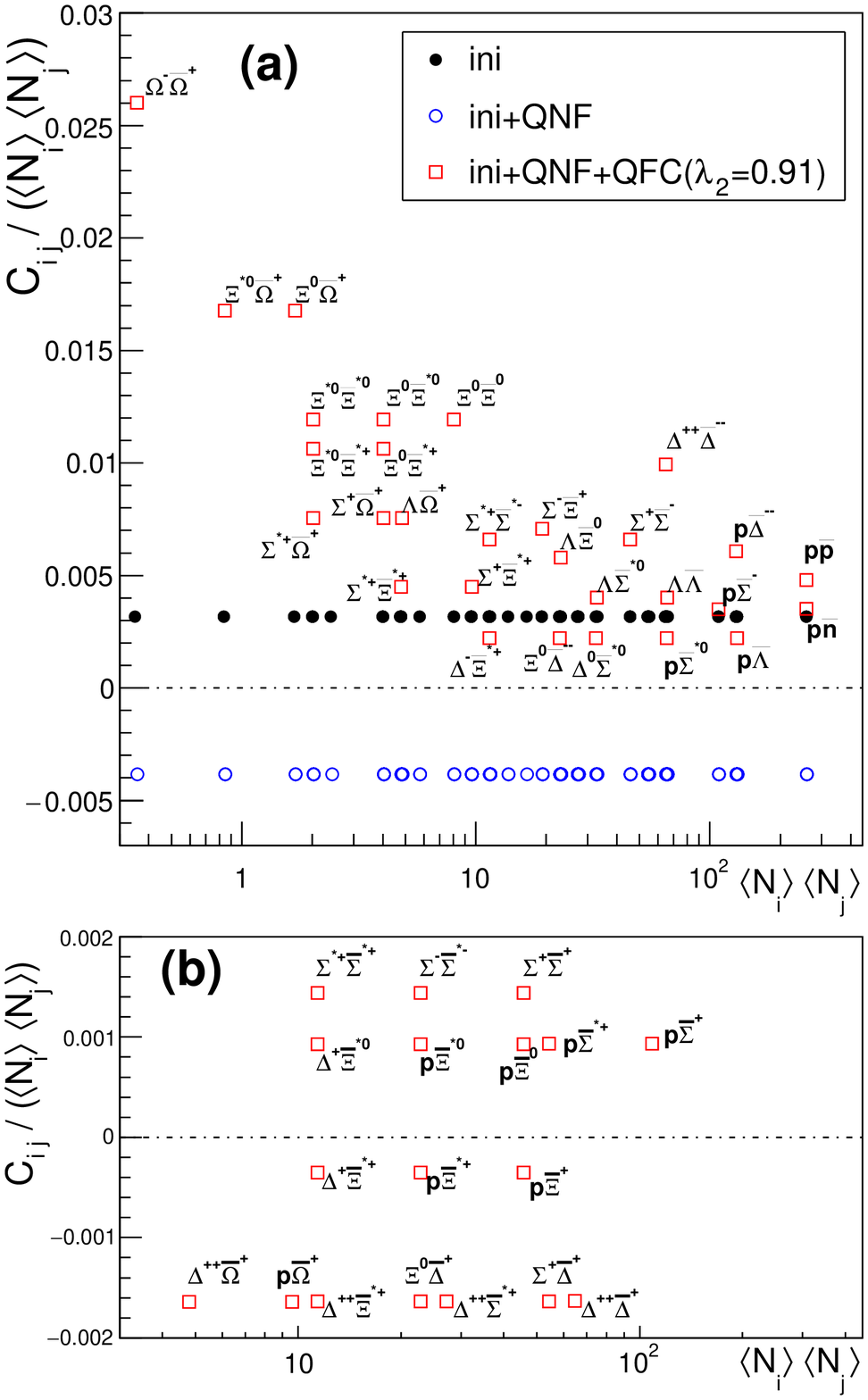}\\
	\caption{(Color online) Baryon-antibaryon multiplicity correlations after considering effects of quark number fluctuations (QNF) and quark flavor conservation (QFC) with parameter $\lambda_2=0.91$. }
  \label{cab_qnfc}
\end{figure}

Fig.~\ref{cab_qnfc} shows various baryon-antibaryon multiplicity correlations after considering effects of quark number fluctuations and correlations.
The system size is taken to be the default value of unit $y_w$.
Solid circles show baryon-antibaryon correlations for the hadronization of the quark system with given quark numbers and antiquark numbers. They exhibit a universal behavior, see Sec.~\ref{subsec_bbar}. After taking into account effects of quark number fluctuations, all baryon-antibaryon correlations, open circles, flip the sign and become a negative and universal value. The negative value means production of baryon and antibaryon is anti-associated. This is because that the increase(decrease) of quark numbers will enhance(suppress) the baryon formation and suppress(enhance) antibaryon formation. It is contrary to the case of two-baryon production discussed in the above subsection.

After further taking into account the flavor conservation of quark numbers with parameter $\lambda_2=0.91$, we get the physical prediction of baryon-antibaryon correlations shown as open squares in Fig.~\ref{cab_qnfc}.  We find that most of baryon-antibaryon correlations return to the positive case which means their production is associated. In particular, hyperon-antihyperon correlations, e.g.,~ $\Omega^{-}\bar{\Omega}^{+}$ and $\Omega^-\bar{\Xi}^{0}$, are much larger than $p\bar{p}$ correlation. This suggests that the strangeness conservation plays an important role in hyperon-antihyperon joint production. Surprisingly, in Fig.~\ref{cab_qnfc} panel (b), some baryon-antibaryon pairs, e.g.,~$p\bar{\Xi}^{+}$, $p\bar{\Omega}^{+}$, have negative values. This is because these baryon-antibaryon pairs do not or less involve the matched $u\bar{u}$, $d\bar{d}$, $s\bar{s}$ pairs and thus flavor conservation less directly constrains their joint production and therefore the effect of quark number fluctuations is dominant.  With small flavor conservation parameter $\lambda_2=0.18$, all baryon-antibaryon correlations tend to zero (with maximum deviation about 0.002) and we do not show them in Fig.~\ref{cab_qnfc} for clarity.

\begin{figure}[!htbp]
\centering
  \includegraphics[width=\linewidth]{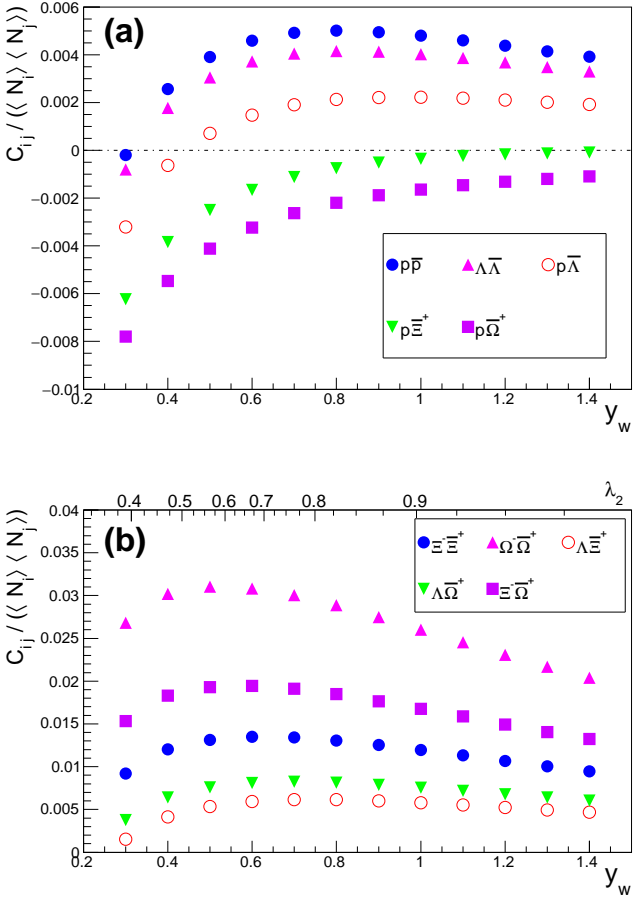}\\
	\caption{(Color online) The rapidity window size dependence of baryon-antibaryon correlations after taking into account quark number fluctuations and correlations just before hadronization. The auxiliary horizontal axis on top of panel (b) shows the corresponding value of flavor conservation parameter $\lambda_2$. }
  \label{cab_stable_qnfc}
\end{figure}

In Fig.~\ref{cab_stable_qnfc}, we show the rapidity window size $y_w$ dependence of some baryon-antibaryon correlations. The relationship between $\lambda_2$ and $y_w$ is the same as that in the above subsection. We observe from panel (a) that $C_{p\bar{\Omega}^{+}}$ is always negative at different $y_w$ and $C_{p\bar{\Xi}^{+}}$ is negative at small $y_w$ and tends to zero with increasing $y_w$ due to the increasing effect of flavor conservation $\lambda_2$. For $p\bar{p}$, $\Lambda\bar{\Lambda}$ correlations in Fig.~\ref{cab_stable_qnfc}(a), $\Xi^{-}\bar{\Xi}^{+}$, $\Omega^{-}\bar{\Omega}^{+}$ and other hyperon-antihyperon correlations in panel (b) that largely involve the matched $u\bar{u}$, $d\bar{d}$, $s\bar{s}$ pairs, they are all positive under the influence of quark flavor conservation. 
We also observe that as $y_w \gtrsim 0.6$, $p\bar{p}$, $\Lambda\bar{\Lambda}$, $\Xi^{-}\bar{\Xi}^{+}$, $\Omega^{-}\bar{\Omega}^{+}$ correlations decrease with the increasing $y_w$, which is because of the increasing quark numbers (or system size). 

\section{Decay effects }
Multiplicity of final baryons observed in experiments usually contains the decay contribution of unstable resonances. In this section, we study the effect of resonance decays on the multiplicity correlations and fluctuations of final stable baryons. We firstly derive formulas of decay influence on stable baryons and then show numerical results of stable baryons $p$, $\Lambda$, $\Xi^{-}$, and $\Omega^{-}$. 

\subsection{formulas of decay effects}
For baryon resonance $i$, its stable daughter baryons are denoted as $a,b,c,\ldots$ with decay branch ratios $\mathcal{D}_{ia}, \mathcal{D}_{ib}, \mathcal{D}_{ic} \ldots$, respectively. $\mathcal{D}_{ij}$ is taken from PDG \cite{pdg2014}. The joint multiplicity distribution of daughter baryons from the parent baryon $i$ of number $N_i$ is taken to be the multinomial distribution $f(\{N_a^i, N_b^i, N_c^i, \ldots\}, N_i, \{\mathcal{D}_{ia}, \mathcal{D}_{ib}, \mathcal{D}_{ic} \ldots\})$, where $N_a^i,N_b^i,N_c^i,\ldots$ denote the numbers of decayed baryons $a,b,c,\ldots$, respectively. Recalling the joint distribution of directly produced baryons in Sec.IV, we write the joint multiplicity distribution of stable baryons 
\begin{widetext}
\begin{equation}
	F(N_a,N_b,N_c,\ldots) = \sum_{\{ N_{h_j} \}}  \mathcal{P}(\{ N_{h_j}\};\{ \langle N_{q_i}\rangle,\langle N_{\bar{q}_i}\rangle \}) \prod_i \Bigg( \sum_{\{N^i\}} f\big(\{N_a^i, N_b^i, N_c^i, \ldots\}, N_i, \{\mathcal{D}_{ia}, \mathcal{D}_{ib}, \mathcal{D}_{ic}, \ldots\}\big)\Bigg) \,\prod_{k=a,b,c,\ldots} \delta_{N_k,\sum_i N_{k}^i},
\end{equation}
\end{widetext}
where index $i$ runs over all kinds of directly produced baryons and $k$ runs over all stable hadrons we study.

The inclusive yield of final-state identified baryons receives the linear superposition of resonance decays,
\begin{equation}
\begin{split}
	&\langle N_{a} \rangle = \sum_{\{N_a,N_b,\ldots\}} N_a \,F(N_a,N_b,N_c,\ldots) \\ 
	&= \sum_{\{ N_{h_j} \}}  \mathcal{P}(\{ N_{h_j}\}) \,  
	\prod_i \Bigg( \sum_{\{N_a^i\}} f\big(N_a^i, N_i, \mathcal{D}_{ia}\big)\Bigg) \, \sum_k N_{a}^k \\
	&=\sum_k \Bigg(	\sum_{\{ N_{h_j} \}}  \mathcal{P}(\{ N_{h_j}\};\{ \langle N_{q_i}\rangle,\langle N_{\bar{q}_i}\rangle \}) \, N_k \mathcal{D}_{ka} \Bigg) \\
	&=\sum_{k} \langle N_k\rangle  \mathcal{D}_{ka} . 
\end{split}
\end{equation}
Note that we have used the abbreviation $\mathcal{P}(\{ N_{h_j}\})\equiv  \mathcal{P}(\{ N_{h_j}\};\{ \langle N_{q_i}\rangle,\langle N_{\bar{q}_i}\rangle \})$ for the joint distribution of directly produced baryons and written $\mathcal{D}_{kk}=1 $ to obtain the compact formulas.  Similarly, we can calculate various moments of multiplicity distributions of stable baryons as 
\begin{equation}
\begin{split}
	&\langle N_{a}^m \rangle = \sum_{\{N_a,N_b,\ldots\}} N_a^m \,F(N_a,N_b,N_c,\ldots) \\ 
	&= \sum_{\{ N_{h_j} \}}  \mathcal{P}(\{ N_{h_j}\}) \,  
	\prod_i \Bigg( \sum_{\{N_a^i\}} f\big(N_a^i, N_i, \mathcal{D}_{ia}\big)\Bigg) \, \Big(\sum_k N_{a}^k \Big)^m, \nonumber
\end{split}
\end{equation}
and finally have  
\begin{equation}
	\sigma_{a}^2 = \sum_{m,n} C_{mn} \, \mathcal{D}_{ma} \mathcal{D}_{na} + \sum_{m} \langle N_m \rangle \mathcal{D}_{ma}(1-\mathcal{D}_{ma} ) ,
\end{equation}

\begin{widetext}
\begin{equation}
	S_a= \frac{1}{	\sigma_{a}^3 } \Bigg( \sum_{k,m,n} C_{kmn} \, \mathcal{D}_{ka} \mathcal{D}_{ma} \mathcal{D}_{na} 
	+ 3\sum_{m,n} C_{mn} \, \mathcal{D}_{ma}(1-\mathcal{D}_{ma})  \mathcal{D}_{na} 
	 + \sum_{m} \langle N_m \rangle \mathcal{D}_{ma}(1-\mathcal{D}_{ma} )(1- 2\mathcal{D}_{ma} ) \Bigg), 
\end{equation}

\begin{equation}
\begin{split}
	K_a + 3 = \frac{1}{	\sigma_a^4 } \Bigg(& \sum_{m,n,k,l} C_{mnkl} \, \mathcal{D}_{ma} \mathcal{D}_{na} \mathcal{D}_{ka} \mathcal{D}_{la} + 6 \sum_{mnk} \Big( C_{mnk} + C_{nk} \langle N_m \rangle \Big) \mathcal{D}_{ma}(1-\mathcal{D}_{ma}) \mathcal{D}_{na} \mathcal{D}_{ka}  \\
	& + 4 \sum_{mn} C_{mn} \mathcal{D}_{ma}(1-\mathcal{D}_{ma}) (1-2\mathcal{D}_{ma})\mathcal{D}_{na}  + 3 \sum_{mn} \Big( C_{mn} + \langle N_m \rangle \langle N_n \rangle \Big) \mathcal{D}_{ma}(1-\mathcal{D}_{ma}) \mathcal{D}_{na}(1-\mathcal{D}_{na}) \\
	& \sum_{m} \langle N_m \rangle \mathcal{D}_{ma}(1-\mathcal{D}_{ma} )\Big[1- 6 \mathcal{D}_{ma}(1- \mathcal{D}_{ma} )\Big] \Bigg).
\end{split}
\end{equation}

The average of the multiplicity product of two stable baryons is evaluated by
\begin{equation}
\begin{split}
	\langle N_{a} N_{b} \rangle &= \sum_{\{N_a,N_b,\ldots\}} N_a N_b \,F(N_a,N_b,N_c,\ldots) 
	= \sum_{\{ N_{h_j} \}}  \mathcal{P}(\{ N_{h_j}\}) \,  
	\prod_i \Bigg( \sum_{\{N_a^i,N_b^i\}} f\big(\{N_a^i,N_b^i\}, N_i, \{\mathcal{D}_{ia}, \mathcal{D}_{ib}\}\big)\Bigg) \, \Big(\sum_m N_{a}^m \Big) \Big(\sum_n N_{b}^n \Big) \\
	& = \sum_{\{ N_{h_j} \}}  \mathcal{P}(\{ N_{h_j}\}) \,  
	\prod_i \Bigg( \sum_{\{N_a^i,N_b^i\}} f\big(\{N_a^i,N_b^i\}, N_i, \{\mathcal{D}_{ia}, \mathcal{D}_{ib}\}\big)\Bigg) \, \Big(\sum_{m\neq n}  N_{a}^m N_b^n + \sum_{m=n} N_a^m N_b^m \Big)  \\
	& = \sum_{m,n} \langle N_m N_n \rangle \mathcal{D}_{ma}  \mathcal{D}_{nb} - \sum_{m} \langle N_m \rangle  \mathcal{D}_{ma}  \mathcal{D}_{mb}.
\end{split}
\label{Nabfinal}
\end{equation}
\end{widetext}
Substituting it into the definition of two-body correlation we get for $a\neq b$
\begin{equation}
	C_{ab} = \sum_{m,n} \big[ C_{mn} -\delta_{mn} \langle N_m \rangle \big] \, \mathcal{D}_{ma} \mathcal{D}_{nb},
\end{equation}
 which receives the coherent superposition of two resonance correlations as well as the anti-association due to the possible same parent resonance. 

 Following the spirit of Eq.~(\ref{Nabfinal}), we obtain the three-body correlation with different species $C_{abc}$ 
\begin{equation}
\begin{split}
	C_{abc} = \sum_{mnk} \Big( & C_{mnk} - (\delta_{mk}+\delta_{nk})C_{mn}  - \delta_{mn}C_{mk} \\
	& + 2\delta_{mn}\delta_{nk} \langle N_m \rangle \Big) \mathcal{D}_{ma} \mathcal{D}_{nb} \mathcal{D}_{kc},  
\end{split}
\end{equation}
and $C_{aac}$ with one identical pair can be obtained by $C_{aac}=\big( C_{abc}\big)_{a=b} + C_{ac}$, 
and the four-body correlation with different species $C_{abcd}$ 
\begin{equation}
\begin{split}
	C_{abcd} = &\sum_{mnkl}  C_{mnkl}  \, \mathcal{D}_{ma} \mathcal{D}_{nb} \mathcal{D}_{kc} \mathcal{D}_{ld}  \\
	&- \sum_{mkl} \Big( C_{mkl} + C_{kl} \langle N_m \rangle  \Big)  \mathcal{D}_{mkl}^{(211)}(a,b,c,d) \\
	&+  \sum_{ml} \Big(C_{ml} + \langle N_m \rangle \langle N_l \rangle \Big) \mathcal{D}_{ml}^{(22)}(a,b,c,d)  \\
	&+ 2 \sum_{ml} C_{ml}\  \mathcal{D}_{ml}^{(31)}(a,b,c,d) \\
	&-6 \sum_{m} \langle N_m \rangle \mathcal{D}_{ma} \mathcal{D}_{nb} \mathcal{D}_{kc} \mathcal{D}_{ld} . 
\end{split}
\end{equation}
Here, $\mathcal{D}_{mkl}^{(211)}(a,b,c,d)= \mathcal{D}_{ma}\mathcal{D}_{mb}\mathcal{D}_{kc}\mathcal{D}_{ld} +\mathcal{D}_{ma}\mathcal{D}_{mc}\mathcal{D}_{kb}\mathcal{D}_{ld}+\mathcal{D}_{ma}\mathcal{D}_{md}\mathcal{D}_{kb}\mathcal{D}_{lc} + \mathcal{D}_{mb}\mathcal{D}_{mc}\mathcal{D}_{ka}\mathcal{D}_{ld}+ \mathcal{D}_{mb}\mathcal{D}_{md}\mathcal{D}_{ka}\mathcal{D}_{lc}+ \mathcal{D}_{mc}\mathcal{D}_{md}\mathcal{D}_{ka}\mathcal{D}_{lb}$ denotes the summation over all possible joint-decay probabilities for three resonances $mkl$ into four stable baryons where the superscript $(211)$ denotes that one of the parent resonances has two decay channels to two different stable baryons, respectively.  Similarly, we have $\mathcal{D}_{ml}^{(31)}(a,b,c,d)= \mathcal{D}_{ma}\mathcal{D}_{mb}\mathcal{D}_{mc}\mathcal{D}_{ld}+ \mathcal{D}_{ma}\mathcal{D}_{mb}\mathcal{D}_{md}\mathcal{D}_{lc}+ \mathcal{D}_{ma}\mathcal{D}_{mc}\mathcal{D}_{md}\mathcal{D}_{lb}+ \mathcal{D}_{mb}\mathcal{D}_{mc}\mathcal{D}_{md}\mathcal{D}_{la}$ and $\mathcal{D}_{ml}^{(22)}(a,b,c,d)= \mathcal{D}_{ma}\mathcal{D}_{mb}\mathcal{D}_{lc}\mathcal{D}_{ld}+ \mathcal{D}_{ma}\mathcal{D}_{mc}\mathcal{D}_{lb}\mathcal{D}_{ld}+ \mathcal{D}_{ma}\mathcal{D}_{md}\mathcal{D}_{lb}\mathcal{D}_{lc}$.  Other four-body correlations of stable baryons with one identical pair, two identical pairs, and three identical species can be obtained as follows
\begin{eqnarray}
	C_{aabd} &=& \Big(C_{abcd}\Big)_{a=c} + C_{abd} + \langle N_a \rangle C_{bd},  \\
	C_{aaab} &=& \Big(C_{abcd}\Big)_{a=c=d} + 3 C_{aab} + \big(3\langle N_a \rangle -2 \big) C_{ab}, \\
	C_{aabb} &=& \Big(C_{abcd}\Big)_{a=c,b=d} + C_{aab} + C_{abb} + \langle N_a \rangle \sigma_b^2   \nonumber \\
	&{}& + \langle N_b \rangle \sigma_a^2 -C_{ab} - \langle N_a \rangle \langle N_b \rangle.
\end{eqnarray}

\subsection{numerical results of stable baryons}
\begin{figure}[!htbp]
\centering
  \includegraphics[width=0.9\linewidth]{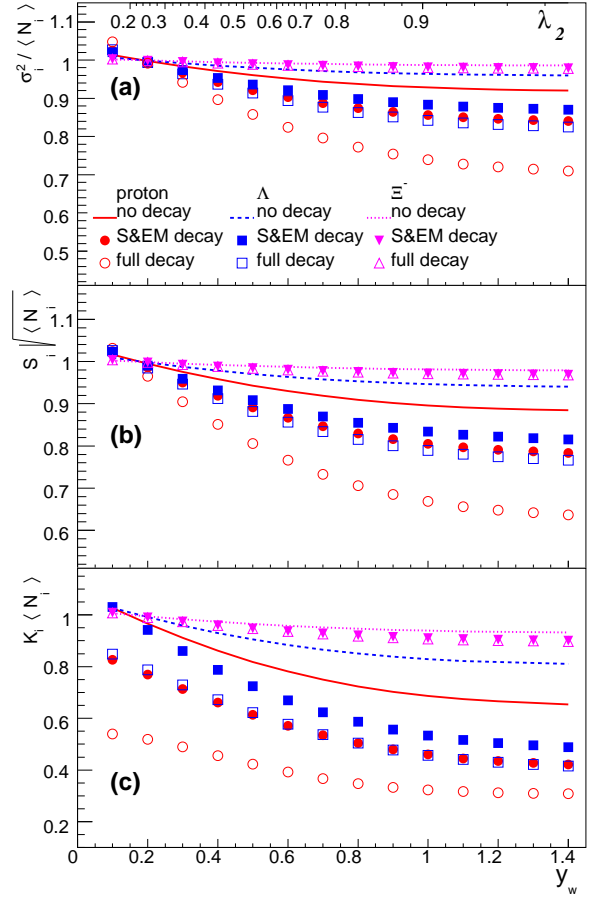}\\
	\caption{(Color online) Moments of final proton, $\Lambda$ and $\Xi^{-}$ at different rapidity window sizes $y_w$. The auxiliary horizontal axis on top of panel (a) shows the corresponding value of flavor conservation parameter $\lambda_2$. Lines show moments of baryons without including resonance decays. Open symbols show results including weak decays, strong decays and electromagnetic decays. Solid symbols show results including only strong and electromagnetic decays. }
  \label{moments_final}
\end{figure}

Fig.~\ref{moments_final} shows multiplicity moments of final proton, $\Lambda$, and $\Xi^{-}$ at different rapidity window sizes. Lines show moments of baryons without including resonance decays. Open symbols show results including weak decays, strong decays and electromagnetic decays. Solid symbols show results including only strong and electromagnetic decays. We can see that due to the large decay contribution to final proton and $\Lambda$, moments of final proton and $\Lambda$, circle and square symbols, are obviously smaller than those of initial ones without including resonance decays, solid and dashed lines, respectively. The decay contribution to $\Xi^{-}$ multiplicity is relatively small, and we see that both weak decays and strong and electromagnetic decays weakly influence moments of $\Xi^{-}$. In contrast to significant $y_w$ dependence of moments of proton and $\Lambda$, moments of $\Xi^{-}$ are only weakly decreased with increasing $y_w$, and the magnitudes are almost one, which is quite close to Poisson distribution. 

\begin{figure}[!htbp]
\centering
  \includegraphics[width=\linewidth]{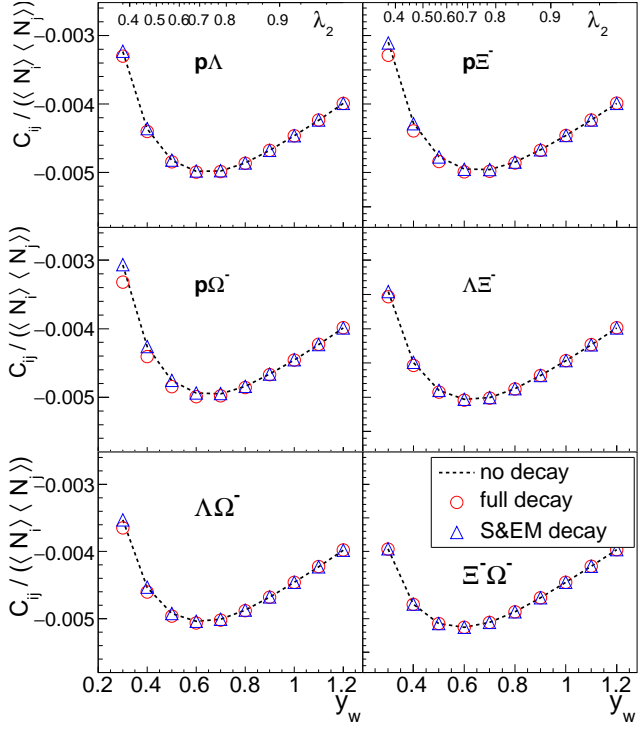}\\
	\caption{(Color online) Two-baryon correlations at different rapidity window sizes $y_w$. The auxiliary horizontal axis on top of panels shows the corresponding value of flavor conservation parameter $\lambda_2$. Dashed lines show baryon-antibaryon correlations without including resonance decays. Open circles show results including weak decays, strong decays and electromagnetic decays. Open up-triangles show results including only strong and electromagnetic decays. }
  \label{c2b_final}
\end{figure}

Fig.~\ref{c2b_final} shows two-baryon correlations of final proton, $\Lambda$, $\Xi^{-}$ and $\Omega^{-}$ at different rapidity window sizes. Surprisingly, we see that they are almost unaffected by resonance decays. However, we emphasize that the almost unchanged quantities are relative correlations $C_{ij}/(\langle N_i \rangle \langle N_j \rangle)$, and for absolute correlations $C_{ij}$ they indeed change a lot. The nonmonotonic dependence of two-baryon correlations on rapidity window size is also a striking behavior for the future experimental measurement. 

\begin{figure*}[!htbp]
\centering
  \includegraphics[width=0.9\linewidth]{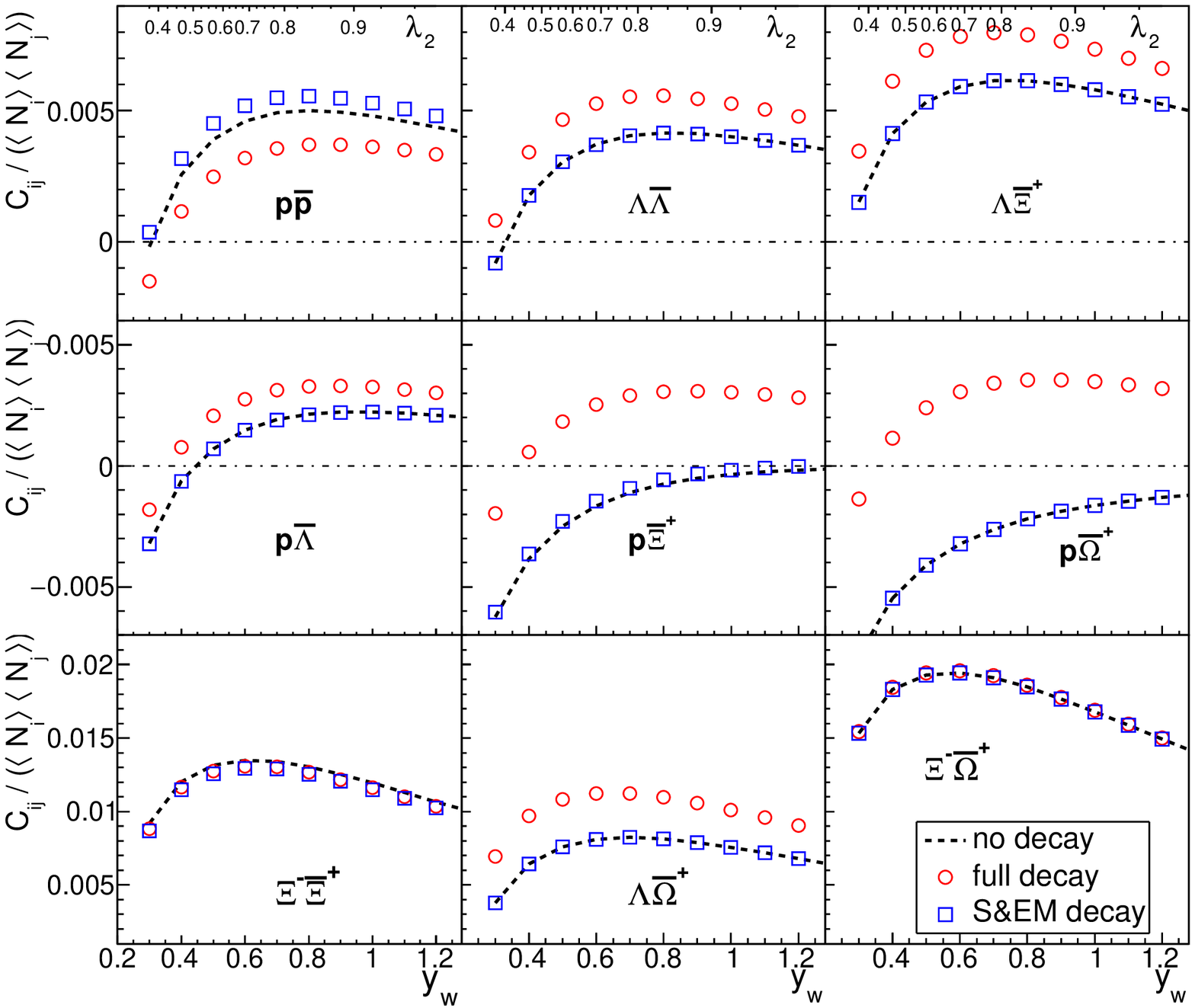}\\
	\caption{(Color online) Baryon-antibaryon correlations at different rapidity window sizes $y_w$. The auxiliary horizontal axis on top of panels shows the corresponding value of flavor conservation parameter $\lambda_2$. Dashed lines show baryon-antibaryon correlations without including resonance decays. Open circles show results including weak decays, strong decays and electromagnetic decays. Open squares show results including only strong and electromagnetic decays.}
  \label{cab_final}
\end{figure*}

Fig.~\ref{cab_final} shows baryon-antibaryon correlations of final proton, $\Lambda$, $\Xi^{-}$ and $\Omega^{-}$ at different rapidity window sizes. Open squares show results including only strong and electromagnetic (S\&EM) decays. Comparing to initial baryon-antibaryon correlations without resonance decays (dashed lines), we can see that all correlations except $p\bar{p}$ are almost unaffected by S\&EM decays. However, for baryon-antibaryon correlations except $\Xi^{-}\bar{\Xi}^{+}$ and $\Xi^{-}\bar{\Omega}^{+}$, after further including weak decays, they (open circles) are significantly changed.
In addition, we observe that final $p\bar{p}$, $p\bar{\Lambda}$, $p\bar{\Xi}^{+}$ and $p\bar{\Omega}^{+}$ with full decay contributions, open circles, have almost the same correlations. This is because that they all reflect such a baryon-antibaryon production association, i.e.,~when an antibaryon either $\bar{p}$, $\bar{\Lambda}$ or $\bar{\Xi}^{+}$ is produced, a baryon of any species (via final proton) should be produced with a certain associated probability to balance the baryon quantum number. 

There are some striking properties in the above decay calculations which are suitable for the future experimental measurement. First, $\Xi^{-}\bar{\Xi}^{+}$ and $\Xi^{-}\bar{\Omega}^{+}$ correlations are almost unaffected by resonance decays.  Second, $p\bar{\Omega}^{+}$ correlation with only S\&EM decays is negative while including weak decays is positive at moderate and large rapidity window sizes. Third, final $p\bar{\Lambda}$ correlation changes the sign around moderate rapidity window size. Fourth, final $p\bar{\Xi}^{+}$ correlation with full decay contribution is positive while including only S\&EM decays it tends to zero at moderate and large $y_w$. 

\section{summary and discussion}
	We have studied dynamical multiplicity fluctuations and correlations of identified baryons and antibaryons produced by the hadronization of bulk quark system in quark combination model. We firstly develop a working model to discuss the most basic dynamics of the quark combination which is necessary to multiplicity study. Then, for the hadronization of quark system with given quark numbers and antiquark numbers, we derive moments (variance, skewness and kurtosis) of multiplicity distributions of produced baryons, two-baryon multiplicity correlations, and baryon-antibaryon multiplicity correlations. We obtain some interesting results about baryon multiplicity as follows. 
\begin{itemize}
			\item[(1)] Multiplicity moments of identified baryons exhibit the behavior of binomial distribution.
			\item[(2)] Anti-association of two-baryon production is mainly determined by the coincide flavors of two baryons.
			\item[(3)] All baryon-antibaryon correlations show a positive and universal magnitude, which suggests that the joint production of baryon and antibaryon is mainly constrained by baryon quantum number conservation in combination.
\end{itemize}	
These properties come from the basic dynamics of the quark combination and, therefore, can be regarded as general features of the quark combination mechanism.
	
	We also take into account correlations and fluctuations of quark numbers and antiquark numbers before hadronization to study their effects on multiple production of baryons and antibaryons. Supposing the weak off-diagonal flavor correlations of quarks and antiquarks, we focus on effects of quark number fluctuations and flavor conservation. In order to relate the experimental measurement at specific rapidity window size $y_w$, we use the charge balance function of thermal particles measured in central Pb+Pb collisions at $\sqrt{s_{NN}}=2.76$ TeV to constrain the flavor conservation at different rapidity window sizes. We calculate moments of inclusive baryon multiplicity, two-baryon multiplicity correlations and baryon-antibaryon correlations at mid-rapidity with unit window size and these at different rapidity window sizes. Comparing with those results directly from the quark combination, after including quark number fluctuations and correlations we find 
\begin{itemize}
		\item[(1)] multiplicity moments of baryons deviate from binomial distribution, and at small flavor conservation parameter we can observe the Poisson statistics; 	
		\item[(2)] all two-baryon correlations at unit rapidity window size tend to be a negative and universal value; 
		\item[(3)] baryon-antibaryon correlations exhibit large species difference. In particular, $C_{p\bar{\Omega}^{+}}$ is negative showing the anti-association between $p$ and $\bar{\Omega}^{+}$ production. At moderate rapidity window size we observe the negative sign of $p\bar{\Xi}^{+}$ correlation but at large window size we observe the vanishing $p\bar{\Xi}^{+}$ correlation. We also observe the sign change of $p\bar{\Lambda}$ correlation at moderate window size. 
\end{itemize}	

	We also study the influence of resonance decays. We separately calculate the above quantities including strong and electromagnetic (S\&EM) decays and those further including weak decays. Our final results of stable baryons $p$, $\Lambda$, $\Xi^{-}$ and $\Omega^-$ show several interesting properties as follows. 
\begin{itemize}
	\item[(1)] Moments of final proton and $\Lambda$ are obviously smaller than those of directly produced baryons. However, the scaled moments of final $\Xi^{-}$ are weakly influenced by resonance decays and are close to Poisson distribution. 
	\item[(2)] Two-baryon correlations are hardly influenced by either S\&EM decays or weak decays. In addition, they are dependent on rapidity window size in a nonmonotonic way. 
	\item[(3)] Effects of resonance decays on baryon-antibaryon correlations are sophisticated.  $\Xi^{-}\bar{\Xi}^{+}$ and $\Xi^{-}\bar{\Omega}^{+}$ correlations are almost unaffected by S\&EM and weak decays.  $p\bar{\Omega}^{+}$ correlation with only S\&EM decays is negative while including weak decays is positive at moderate and large rapidity window sizes. $p\bar{\Lambda}$ correlation changes the sign around moderate rapidity window size. 
\end{itemize}
They are striking phenomena which are suitable for the future experimental measurement. 

Some discussions related to experimental observation at finite rapidity window size are in order. In Sec.~III and IV, we choose a quark system of specific size which corresponds to a specific rapidity window of the quark system produced in relativistic heavy ion collisions.  Here we do not consider the possible rapidity shift between (anti-)quarks and the formed (anti-)baryon, which may lead to the produced baryons to fly off the studied window and baryons produced in other region to fly into this window. However, the effect of rapidity shift in combination is quite small because of the following two reasons. First, there is small discrepancy between the total mass of three quarks and the mass of the formed baryon. Note that we usually use the constituent quark mass in QCM, i.e.,~$m_u \sim 330$ MeV and $m_s \sim 500 $ MeV. Therefore, there is no large rapidity shift in combination due to the mass (or energy) mismatch between three neighboring quarks in phase space and the baryon they form. Second, we apply the quark combination rule as explained in Sec.~II to longitudinal rapidity direction to solve the unitary issue which is necessary for multiplicity study. This approach has reproduced experimental data of rapidity distributions of identified hadrons in relativistic heavy ion collisions at different collisional energies. The rapidity interval between neighboring quarks is only the order of $10^{-3}$ due to the high quark number density $dN/dy\sim 10^3$ in ultra-relativistic heavy ion collisions. Therefore, rapidity shift in baryon production is quite small and it hardly influences results in this work. In Sec.~V, we also neglect the rapidity shift in resonance decays. Because the rapidity shift in baryon decays is small ($\lesssim 0.1$), its influence is also expected to be small. 

\section*{Acknowledgments}
	The authors thank Shu-qing Li for helpful discussions. The work is supported by the National Natural Science Foundation of China under grants Nos. 11305076, 11575100, 11505104 and 11675091.

\appendix*
\section{derivation of Eqs.~(\ref{vbi_qnfc}-\ref{kbi_qnfc})}
Applying Eq.~(\ref{enav}) and substituting the following expansion 
\begin{equation}
\begin{split}
	\langle N^{m}_{\alpha} N^{n}_{\beta} \rangle =\  &\overline{N^{m}_{\alpha} N^{n}_{\beta}} + \frac{1}{2}\sum_{f_1 f_2} \partial_{12}\overline{N^{m}_{\alpha} N^{n}_{\beta}}\,C_{f_1 f_2} \\
	&+ \frac{1}{3!}\sum_{f_1 f_2 f_3} \partial_{123}\overline{N^{m}_{\alpha} N^{n}_{\beta}}\, C_{f_1 f_2 f_3} \\
	&+ \frac{1}{4!}\sum_{f_1 f_2 f_3 f_4} \partial_{1234}\overline{N^{m}_{\alpha} N^{n}_{\beta}}\, C_{f_1 f_2 f_3 f_4} 
\end{split}
\end{equation}
into the definition of multiplicity moments 
\begin{eqnarray}
	\sigma^2_{\alpha} &=& \langle N^2_{\alpha}\rangle - \langle N_{\alpha} \rangle^2, \nonumber \\
	\langle \delta N^3_{\alpha} \rangle &=& \langle N^3_{\alpha} \rangle - 3\langle N_{\alpha} \rangle \sigma^2_{\alpha} - \langle N_{\alpha} \rangle^3, \label{A2}  \\
	\langle \delta N^4_{\alpha} \rangle &=& \langle N^4_{\alpha} \rangle - 4\langle \delta N^3_{\alpha} \rangle \langle N_{\alpha} \rangle - 6\langle N_{\alpha} \rangle^2 \sigma^2_{\alpha} - \langle N_{\alpha} \rangle^4,  \nonumber
\end{eqnarray}
and two-body multiplicity correlation
\begin{equation}
	C_{\alpha \beta} = \langle N_{\alpha} N_{\beta} \rangle - \langle N_{\alpha} \rangle \langle N_{\beta} \rangle, 
	\label{A3}
\end{equation}
 we can get the expressions of Eqs.~(\ref{vbi_qnfc}-\ref{kbi_qnfc}) up to two-body quark correlations. The complete expansions of Eqs.~(\ref{A2}) and (\ref{A3}) up to four-body quark correlations are too long to be shown. 
In addition, direct calculations according to Eqs.~(\ref{A2}) and (\ref{A3}) are numerically convenient.

\end{document}